\documentclass{aastex}
\usepackage{emulateapj5}
\usepackage{amsmath}

\newcommand{\HI}{H$\,${\sc i}}

\newcommand{\ea}{{\rm et\,al.}}
\newcommand{\kms}{km\,s$^{-1}$}
\newcommand{\holmi}{Ho\,I}
\newcommand{\Ha}{\mbox{H$\alpha$}}
\newcommand{\msol}{${\cal M}_\sun$}
\defcitealias{sch98}{Schlegel et al. 1998}

\slugcomment{accepted for publication in the Astronomical Journal (AJ)}

\shorttitle{Blow--out in {\holmi}}
\shortauthors{Ott {\ea}}

\begin{document}

\title{Evidence for Blow--out in the Low--mass Dwarf Galaxy Holmberg\,I}

\author{J\"urgen Ott\altaffilmark{1}}
\affil{Radioastronomisches Institut der Universit\"at Bonn, 
       Auf dem H\"ugel 71, 53121 Bonn, Germany\\jott@astro.uni-bonn.de}
\author{Fabian Walter\altaffilmark{1}}
\affil{California Institute of Technology, Owens Valley Radio Observatory, Astronomy Department 105--24, 
       Pasadena, CA 91125, USA\\fw@astro.caltech.edu}
\author{Elias Brinks}
\affil{Departamento de Astronom\'{\i}a, Universidad de Guanajuato, Apartado Postal 144, Guanajuato, Gto 36000, 
       Mexico\\ebrinks@astro.ugto.mx}
\author{Schuyler D. Van Dyk}
\affil{Infrared Processing and Analysis Center/California Institute of Technology, 
       Mail Code~100--22, Pasadena, CA 91125, USA\\vandyk@ipac.caltech.edu}
\author{Boris Dirsch}
\affil{Universidad de Concepci\'on, Departamento de F\'{\i}sica, Casilla 160--C, Concepci\'on, Chile
\\bdirsch@cepheid.cfm.udec.cl}
\and
\author{Ulrich Klein}
\affil{Radioastronomisches Institut der Universit\"at Bonn, Auf dem H\"ugel 71, 53121 Bonn, Germany\\uklein@astro.uni-bonn.de}
\altaffiltext{1}{Visiting Astronomer, German--Spanish Astronomical Centre, Calar Alto, 
operated by the Max--Planck--Institute for Astronomy, Heidelberg, jointly with the Spanish National Commission for Astronomy.}

\begin{abstract}
  
  We present radio and optical observations of Holmberg\,I ({\holmi}),
  a member of the M\,81 group of galaxies (distance $\sim
  3.6$\,Mpc). {\holmi} is a low--mass, low surface--brightness dwarf
  galaxy. High--resolution multi--array Very Large Array observations
  in the line of neutral hydrogen (\ion{H}{1}) reveal a supergiant
  shell (diameter 1.7\,kpc) which covers about half the optical extent
  of {\holmi} and which comprises 75\% of the total \ion{H}{1} content
  (total \ion{H}{1} mass: $1.1 \times 10^8\,{\cal M}_\sun $). We
  estimate the scaleheight of the \ion{H}{1} layer to be 250\,pc
  $\lesssim h \lesssim 550$\,pc. We set a tentative upper limit to the
  dark matter content of $\lesssim 3.1 \times 10^{8}\,{\cal
  M}_{\sun}$. The \ion{H}{1} data are complemented by deep, optical
  UBV(RI)$_c$ observations and narrow band {\Ha} imaging obtained at
  the Calar Alto 2.2\,m telescope. We find ${\cal M}_{\rm HI}$/$L_{\rm
  B} = 1.1\,{\cal M}_\sun / L_{{\rm B}_\sun}$. The total visible
  (stars plus gas) mass of {\holmi} adds up to $2.4\times10^8 \,{\cal
  M}_\sun$. This leads to a total mass of $\lesssim 5.5 \times
  10^{8}\,{\cal M}_{\sun}$ and an inclination for {\holmi} of
  $10{\degr}\lesssim i \lesssim 14{\degr}$.
  
  The origin of {\holmi}'s peculiar \ion{H}{1} morphology is discussed
  in terms of a supergiant shell created by strong stellar winds and
  supernova explosions. We estimate that the energy deposited falls in
  the range of $1.2 \times 10^{53}$\,erg $\lesssim E \lesssim
  2.6\times10^{53}$\,erg (equivalent to 120--260 type II SN
  explosions). From a comparison with isochrones as well as from
  dynamical modeling based on the \ion{H}{1} data we derive an age for
  the supergiant \ion{H}{1} shell of $\sim 80\pm20$\,Myr. The
  morphological center of {\holmi} (i.e., the center of the ring) is
  offset by 0.75\,kpc with respect to the dynamical center. Within
  the interior of the shell the light distribution is exponential with
  a rather shallow gradient and blue optical colors. Beyond a radius
  corresponding to an \ion{H}{1} column density of
  $\sim10^{21}$\,cm$^{-2}$, the putative star formation threshold, the
  disk becomes considerably redder and the slope for the exponential
  light distributions steepens. We attribute this to a uniform star
  formation activity in the recent past within the central 2~kpc of
  {\holmi}. Color--magnitude diagrams based on our CCD data show that
  the youngest stars, with ages of 15--30\,Myr, are situated along the
  inside of the rim of the giant \ion{H}{1} shell which is where we
  also find some faint \ion{H}{2} regions. It is speculated that these
  stars are the result of secondary star formation on the rim of the
  shell. Based on the global morphology and velocity dispersion as
  well as the location of the \ion{H}{2} regions we find evidence for
  ram pressure within the M\,81 group. Finally, we discuss the
  likelihood of {\holmi} having lost some of its interstellar material
  to the intergalactic medium (``blow--out'' scenario).

\end{abstract}

\keywords{galaxies: individual (Holmberg I, UGC 5139, DDO 63) ---
galaxies: irregular --- galaxies: dwarf --- galaxies: photometry ---
ISM: bubbles --- ISM: HI}



\section{Introduction}
\label{intro}

Star formation (SF), and subsequent evolution of the most massive
stars, is believed to play a dominant role in shaping the interstellar
medium (ISM) of galaxies. Stars with masses larger than 8\,{\msol}
first produce wind--blown bubbles within which core--collapse
supernovae (SNe) occur, typically on time scales of 0.3 -- $4 \times
10^7$ years \citep{lei99}. Each of these type II SNe releases a total
energy of $\sim (2-3)\times 10^{53}$\,erg, mostly in the form of
neutrinos of all flavors \citep[for a review see][]{bur00}. Only about
1\% ($\sim 10^{51}$\,erg) of this energy is deposited as kinetic
energy in the environment. It is this fraction which ultimately
creates features in the ISM with typical dimensions of tens of parsecs
known as ``bubbles'' or ``shells,'' depending whether one concentrates
on the inferred three--dimensional morphology or the observed
two--dimensional shape. As SF tends to occur in associations and
clusters of stars one finds that the combined effects of stellar winds
and SNe of an ensemble of high--mass stars can create larger features
such as ``superbubbles'' or ``supergiant shells'' with radii of up to
more than 1\,kpc. Recent models of an evolving young association with
a total stellar mass of $10^4$\,{\msol} predict a total kinetic energy
input of $\sim 10^{53}$\,erg \citep{lei99}. Once no further energy
input is provided, i.e., after the lowest mass stars which end their
lives as supernovae have disappeared, the expansion velocity of the
SN--driven bubbles and shells will decrease to values similar to the
velocity dispersion of the surrounding ISM.  Eventually they will
stall. This state of affairs is reached sooner for those bubbles which
break--out of the disk of their host, leaving a morphology of a ring
rather than a bubble.

The expanding shock front around each bubble piles up the surrounding
ISM on its rim \citep[see, e.g., the reviews by][and references
therein]{ten88,bri98} and, if densities and temperatures are
appropriate, new SF can occur. This process, known as ``propagating
star formation'' \citep[e.g.,][]{mue76,ger78,elm94}, extends the time
span for energy input and can lead to even larger shells, provided the
bubble has not broken out of the disk. The recurrent deposition of
such huge amounts of energy and their transformation into kinetic
energy leads to a heating of the ISM, providing a natural energy
source for the observed velocity dispersion of the (neutral) gas
component.

It should be noted, though, that observational evidence for a stellar
origin of superbubbles or supergiant shells is scarce. Only in very
few cases are stellar clusters found within giant
\ion{H}{1} holes \citep{ste00}. In most cases these structures do
not seem to host remnant stellar clusters, as was pointed out by
\citet{rho99}.

Observationally, \citet{hei79,hei84} was the first to study the
\ion{H}{1} morphology and dynamics of bubbles and shells
in the ISM of the Milky Way. This was followed by an analogous study
of the ISM in M\,31 by \citet{bri86}. They found structures up to
700\,pc in diameter. Considerably larger shells have been found in
dwarf irregular galaxies (dIrr). This can be understood as follows:
the low gravitational potential binds the ISM less tightly to the
disk. As mentioned by \citet{wal99}, the \ion{H}{1} velocity
dispersion is approximately the same in dwarf galaxies as in larger,
spiral galaxies ($\sim 6-9$\,{\kms}). Hence the scaleheight in dIrrs
is larger than in spirals. This holds true in absolute as well as in
relative terms. As a result, the volume density of the ISM is lower
compared to grand--design spiral galaxies, resulting in a lower
resistance against the kinetic energy input of SF regions. Secondly,
the shells can evolve over a longer period. This is due to the larger
scaleheight which delays and possibly even prevents
break--out. Moreover, since dIrrs show mostly solid body rotation, there
is little shear that may destroy shells once they have formed.
Examples of objects which show large \ion{H}{1} shell radii are
Holmberg\,II \citep{puc92}, IC\,2574 \citep{wal99}, IC\,10
\citep{wil98}, DDO\,47 \citep{wal01}, NGC\,6822 \citep{blo00}, 
and the Magellanic Clouds \citep{sta97,kim99}.

To first order at least, a region of massive SF is oblivious to its
environment, dumping the same amount of energy into the ISM whether it
finds itself in a large spiral galaxy or in a dwarf. Therefore, for
progressively lower overall galaxy mass, the impact of SF and its
aftermath becomes proportionally more and more important. Theoretical
models, e.g., by \citet{dey94}, \citet{mac99}, and \citet{fer00},
predict the loss of metal--enriched interstellar material associated
with low--mass dwarf galaxies (${\cal M}_{visible}\lesssim
10^{9}\,{\cal M}_{\sun}$) in the form of shells breaking out of the
disk (``blow--out'' scenario). As dIrrs are dominant in number in the
universe and are considered to be the building blocks of
larger galaxies in ``bottom--up'' scenarios of galaxy formation, this
may have implications with respect to the enrichment of the
intergalactic medium (IGM) at larger look--back times.

In some low--mass dwarf galaxies the ISM shows an almost perfect ring
feature in \ion{H}{1}, more or less centered on the optical
galaxy. According to the models discussed above this may be indicative
of a recent starburst in the center of the galaxy which created the
cavity. In these galaxies, the total mass may just have been high
enough to retain the expanding material. Objects with even lower mass
(${\cal M}_{visible}\lesssim 10^{6}\,{\cal M}_{\sun}$) would have
suffered ``blow--away'' \citep{fer00} of the entire ISM by a similar
event. Examples of gaseous rings in dwarf irregular galaxies which
suggest blow--out are Leo\,A \citep{you96}, Sag\,DIG \citep{you97},
M\,81\,dw\,A \citep{wes93}, Holmberg\,I \citep[{\holmi}, ][]{tul78},
Cas\,1 (Huchtmeier, priv.\ comm.), and Sextans\,A \citep{ski88,dyk98}.

Multiwavelength studies, as presented in this paper, are indispensable
if we wish to understand these unique systems. The particular object
under study here is Holmberg I ({\holmi}), a dwarf irregular galaxy in
the nearby M\,81 group of galaxies. This object was detected by
\citet{hol50} who optically searched for members of the M\,81 group.
Since then, relatively few papers have been dedicated to detailed
studies of {\holmi}. \citet{san74} and \citet{tik92} performed
photographic photometry. Both papers concentrate on obtaining a
distance measurement using the brightest stars method. \citet{hoe84},
using early CCD imaging, studied the stellar population of {\holmi} in
the Gunn $gri$ system. Radial surface brightness profiles have been
obtained by \citet{bre98} and \citet{mak99}, both of which observed
{\holmi} in the context of larger surveys of dwarf galaxies. The
luminosity and size of \ion{H}{2} regions, as well as the metallicity
of {\holmi}, have been studied in detail by \citet{mil94,mil96}.
Westerbork \ion{H}{1} radio synthesis observations were presented by
\citet{tul78} whereas \citet{puc93} presented a preliminary map of the
VLA data which are reported in full in this paper.

In this paper we aim to present a consistent picture of Holmberg\,I,
dealing with the properties and evolution of the stars and gas. In
particular, we investigate the possible influence of the stellar
population on the overall gas morphology and the star formation
history of this galaxy. In Sect.\,\ref{obs} we present optical and
\ion{H}{1} observations and describe the results in
Sect.\,\ref{result}. The wealth of data is discussed in
Sect.\,\ref{discuss} and our conclusions are summarized in
Sect.\,\ref{summary}.


\section{Observations and data reduction}
\label{obs}

In this paper we use radio and optical data to study the properties of
the dIrr galaxy Holmberg\,I ({\holmi}, also known as UGC\,5139 or
DDO\,63). We combine maps in the 21--cm line of atomic neutral
hydrogen (\ion{H}{1}) with UBV(RI)$_c$ and {\Ha} observations in the
optical regime in order to study the gas and the stars, respectively.
As {\holmi} is a member of the M\,81 group we assume it to be at the
same distance as M\,81, $3.63 \pm 0.34$\,Mpc which corresponds to
$m-M=27.80$\,mag \citep{fre94}. This is not too different from a
direct measurement
for {\holmi} itself, based on the brightest stars method utilizing
photographic plates \citep{san74}: $m-M=27.63$\,mag. However, using
the same technique, \citet{tik92} found a distance module of
$m-M=29.11$\,mag, which deviates substantially from the others.


\subsection{Optical observations}
\label{optobs}
Optical images were obtained at the Calar Alto\footnote{The Calar Alto
Observatory is operated by the Max--Planck--Institut f\"ur Astronomie
(Heidelberg) jointly with the Spanish Comisi\'on Nacional de
Astronom\'{\i}a.} 2.2--m telescope. The instrument used was a
combination of the focal reducer CAFOS (f/4.2) and the CCD Site\#1d.
This CCD has a pixel size of $0\farcs53\times0\farcs53$, a readout
noise of 5.06 e$^{-}$, and a gain of 2.3 e$^{-}$/ADU. The Johnson
filters B, V, and the Johnson--Cousins filter R$_c$ were used with
integration times of 60\,min, 45\,min, and 35\,min respectively. These
observations were performed in 1999 January under seeing conditions of
$\sim 1\farcs4$. In 2000 January we completed the UBV(RI)$_c$ imaging
with the U (100\,min integration time) and I$_c$ (90\,min)
filters. The seeing was again $\sim 1\farcs4$. Unfortunately the
I$_c$ exposure suffers from fringing at the $\approx 2.5\%$ level of
the global sky background. This results in the loss of faint
structures and a reduced limiting magnitude. Additionally, an
H$\alpha$ image (1999 January) with 60\,min integration time was
obtained. Due to the bandwidth of the filter (12\,nm), this image
still contains some [\ion{N}{2}] emission. All images were
bias--subtracted using an average value of the overscan
region. Flatfield exposures were produced by taking the median of
different skyflats of the same night. All reduction steps were
performed using the IRAF\footnote{IRAF is distributed by National
Optical Astronomy Observatories, which are operated by the Association
of Universities for Research in Astronomy, Inc., under cooperative
agreement with the National Science Foundation.}
\citep{tod93} software package.

All observations were split into three or more exposures, allowing a
straightforward elimination of cosmic ray pixels and other image
defects by applying the CCD clipping algorithm {\sc ccdclip}.
Calibration was performed with $\sim 15$ Landolt standard stars
\citep{lan92} solving for the first--order zero point, the
airmass, and color terms. Fitting of the calibration coefficients was
done using the task {\sc fitparams}. The systematic error in the
calibration is estimated to be 0.05\,mag in B, V, R$_c$, and 0.1\,mag
in U and I$_c$.

For surface brightness measurements, notoriously bright foreground
stars were replaced by a Gaussian background distribution, equal in
mean and scatter to the ambient brightness. We applied a median
filter of $3\farcs2$ in size in order to achieve smoothness. The data
were corrected for Galactic extinction E$_{\rm B-V}$=0.048\,mag at the
position of {\holmi} \citep{sch98} which was converted to the
Johnson(--Cousins) bandpasses following \citet{car98}. In the case of
surface brightness measurements one is interested in diffuse emission
and cannot apply the aperture used for single stars. 
Therefore we had to compare the total flux $F_{tot}$ of a point
source with the flux $F_{aper}$ within the aperture used. To determine
$F_{tot}$, we opened the aperture for several stars in steps of
0\farcs5 and measured the corresponding flux until reaching a constant
value in the limit. We assumed this limit to be $F_{tot}$. The
difference $F_{tot}-F_{aper}$ leads to corrections of 0.06\,mag in B,
V, R$_c$, 0.11 in U and 0.30 in I$_c$.

Crowded field photometry was performed with {\sc DAOPHOT\,II}
\citep{ste87,ste90} as implemented in {\sc IRAF}. We used {\sc
 daofind} to identify candidates of unresolved sources. Subsequently
we determined a PSF magnitude by a manual selection of $\sim 20$
sufficiently bright point sources in the field. We rejected remaining
cosmic ray pixels, slightly resolved background galaxies and image
artifacts by judiciously employing the $sharpness$ and $\chi$
parameters that were determined during the PSF fitting. The
photometric errors of the remaining point sources are displayed in
Fig.\,\ref{formalerrors} (because of the residual fringing we refrained
from attempting any stellar crowded field photometry in the I$_c$ band).


\placefigure{formalerrors}


\subsection{{\ion{H}{1}} observations}
\label{HI}

{\holmi} was observed with the NRAO Very Large Array\footnote{The Very
  Large Array (VLA) is a telescope of the National Radio Astronomy
  Observatory (NRAO) which is operated by Associated Universities,
  Inc., under a cooperative agreement with the National Science
  Foundation.} (VLA) in B--, C--, and D--configuration (1990 July,
1990 December, and 1991 March). The parameters of the observations are
listed in Table\,\ref{tabhiprop}. For the data reduction we used two
software packages: calibration was done with AIPS\footnote{The
  Astronomical Image Processing System (AIPS) was developed by the
  NRAO.} and imaging with MIRIAD\footnote{MIRIAD was developed by the
  ATNF (Australia Telescope National Facility) which is part of the
  CSIRO (Commonwealth Scientific and Industrial Research
  Organization).} \citep{sau95}. To visualize the data, we made
extensive use of the ATNF package KARMA \citep{goo95}.


\placetable{tabhiprop}


To start with, unreliable visibilities were flagged by visual
inspection of the raw {\it uv}--data after which we applied flux,
complex gain, and bandpass calibrations using the tasks {\sc setjy,
 getjy, calib {\rm and} bpass}. To extract the 21\,cm line emission,
a linear interpolation of the continuum in the {\it uv}--plane was
applied by taking the average of line--free channels on each side of
the passband. This interpolation was subsequently subtracted from all
visibilities (task {\sc uvlin}). As a final step in AIPS, the
calibrated datasets from all configurations were combined with {\sc
 dbcon}.

In the imaging process the {\it uv}--data were Fourier transformed
using both ``natural'' and ``uniform'' weighting (task {\sc invert}
within MIRIAD). As both weighting schemes have some mutually exclusive
advantages, it is desirable to find the optimum between low noise and
high resolution. Therefore the visibilities were also weighted with
the ``robust''--weighting algorithm, as originally developed by
\citet{bri95}. Deconvolution of the resulting cube was subsequently
performed by applying the Clark {\sc Clean}--algorithm \citep{cla80}
(tasks {\sc clean} and {\sc restor}). The ``robust'' data cube has a
synthesized beam of 8\farcs2 $\times$ 7\farcs0 in size (approximately
140\,pc $\times$ 120\,pc) with a position angle of $-73^\circ$ and an
rms noise per channel map of 1.4\,mJy\,beam$^{-1}$ (equivalent to a brightness
temperature of 14.8\,K) which corresponds to a column
density of 7$\times 10^{19}$\,cm$^{-2}$.

In order to decide which emission at a faint level is real in the data
cube, we produced a ``master cube'' where all noise dominated positions
were replaced by ``blank'' values. To do this, we first smoothed the
data cube to 20{\arcsec} and blanked out all positions where the
convolved signal drops to below 2.5$\sigma$ rms. Secondly, the resulting
cube was examined by eye for emission in each channel map. Only
regions with signal in three or more consecutive channels were
considered to contain genuine emission. The same areas which were
blanked in the ``master cube'' were blanked as well in the ``robust''
and ``natural''--weighted cube. These blanked cubes were subsequently
used for the calculation of the flux and the moment maps, i.e., the
velocity--integrated \ion{H}{1} distribution, the velocity field, and
the \ion{H}{1} velocity dispersion map (task {\sc moment} in
MIRIAD).

The data cube thus obtained is a combination of a {\sc clean}ed image
and a residual dirty image. Whereas the {\sc clean}--beam is well
defined, the effect of the residual dirty beam on the output maps is
{\em a priori} unknown. This makes the determination of the flux level
in each channel map uncertain and a correction needs to be applied.
We used the method proposed by \citet{joe95} \citep[see
also][]{wal99}. The total flux $S$ can be computed by determining the
fluxes of the dirty data cube $D$, the residual cube $R$, and the {\sc
clean} components convolved with the {\sc clean} beam $C$ via the
equation
\begin{equation}
S\,=\,\frac{C\,\times\,D}{D\,-\,R}.
\end{equation}
We note that we would have overestimated our \ion{H}{1} flux by $\sim 70$\,\%
if we had not applied this correction.


\section{Results}
\label{result}


\subsection{Global Optical Properties}
\label{optprop}

Figure\,\ref{holmi_rgb} shows a true color image of \holmi\ based on a
combination of our UBVR$_c$ CCD frames. One can clearly discern an
underlying population of red objects with superimposed large
collections of blue and presumably young stars (especially towards the
southern part of the galaxy).


\placefigure{holmi_rgb}


Before attempting photometry on individual stars or star clusters we
determined apparent and absolute integrated magnitudes for {\holmi}
(see Table\,\ref{opttable}). Our limiting surface brightness
magnitudes are approximately 26\,mag/$\sq {\arcsec}$ (U), 27\,mag/$\sq
{\arcsec}$ (B), 26\,mag/$\sq {\arcsec}$ (V), 26\,mag/$\sq {\arcsec}$
(R$_c$), and 23\,mag/$\sq {\arcsec}$ (I$_c$).

\noindent Taking 
\begin{equation} 
L_{\rm B} = 10^{-0.4(M_{\rm B} - M_{{\rm B}_{\sun}})}L_{{\rm B}_\sun} 
\end{equation} 

\noindent and a solar absolute blue magnitude of M$_{{\rm B}_\sun} = 5.50$\,mag
\citep{lang92} this leads to $L_{\rm B} = 1.0 \times 10^8 L_{{\rm
  B}_{\sun}}$. This value is corrected for Galactic foreground
extinction corresponding to E$_{\rm B-V}$=0.048\,mag \citepalias{sch98} but
not for any (unknown) internal extinction within {\holmi}. However,
since \citet{mil96} derived a metallicity of 12+log(O/H)=7.7
($\approx$8\% solar) and extinction is correlated with metallicity, we
expect the internal extinction to be negligible.


\placetable{opttable}


\subsection{\ion{H}{1} morphology and dynamics}


\subsubsection{\ion{H}{1} distribution}
\label{hidistr}

Fig.\,\ref{channmap} shows the ``robust'' weighted \ion{H}{1} channel
maps. The beam size is indicated in the lower left--hand of each
panel. It is only a few times larger than the seeing of our optical
observations. The region to the north--west of the center appears in
nearly all channel maps, suggesting a large velocity dispersion. The
area near the central position shows virtually no emission throughout
the dataset. The total \ion{H}{1} diameter of {\holmi} is $\sim
5.8$\,kpc.


\placefigure{channmap}
\notetoeditor{The Fig.\,\ref{channmap} consists of two images (fig3a and fig3b)}

     
We used the most sensitive ``natural''--weighted cube for the
determination of the spectrum and the flux. The \ion{H}{1} spectrum
of {\holmi} is shown in Fig.\,\ref{spectrum}. Its shape is remarkably
well described by a Gaussian with a FWHM of 27.1\,{\kms} and a central
velocity of $\sim 140$\,{\kms}. Although it is not clear, a priori, why
this should be so, this was noticed earlier by \citet{lo93} who, at
somewhat lower velocity resolution, found similar \ion{H}{1} spectra
in nine intrinsically faint dwarf galaxies. \citet{sti99} also
mentioned that dwarf galaxies with a central \ion{H}{1} hole show a
single peak in their spectra. We tried to model the \ion{H}{1}
distribution of {\holmi} with the task {\sc galmod}, which is part of
the software package GIPSY\footnote{The ``Groningen Image Processing
System''}. We started with the observed properties of
{\holmi}. Changing parameters like the rotation curve and/or the
\ion{H}{1} column density distribution while keeping the inclination
and the dispersion fixed, led to a bimodal \ion{H}{1} spectrum. On the
other hand, varying the dispersion and/or the inclination,
and leaving the column density distribution and the rotation curve
untouched, kept the Gaussian shape of the spectrum.
     

\placefigure{spectrum}

 
Integrating the flux over all channels yields $36.0 \pm
4.0$\,Jy\,{\kms}, corresponding to a total \ion{H}{1} mass of
$1.1\times 10^8\,{\cal M}_{\sun}$. This result places {\holmi} in the
intermediate to low--mass range of that found for dwarf irregular
galaxies. Our values compare well with earlier single--dish
measurements by \citet{dic78} who found 39.7\,Jy\,{\kms}. This
indicates that we hardly miss any extended emission in our
interferometric observations. Note that the value quoted by
\citet{all79} of 49.0\,Jy\,{\kms} is off by a fairly large margin. We
do not have an explanation for this difference other than possible
confusion with ambient neutral (Galactic) gas within their large
beam. The observed \ion{H}{1} mass leads to an ${\cal M}_{\rm
HI}$/$L_{\rm B}$ ratio of 1.1\,${\cal M}_\sun / L_{{\rm B}_\sun}$.

A prominent depletion of atomic gas is seen in the center of the
integrated \ion{H}{1} map (Fig.\,\ref{mom0}), while the bulk of the
emission is found in a ring. The highest column densities, $2.0\times
10^{21}$\,cm$^{-2}$, are encountered towards the south--east, where
the column density is marginally higher than the average value in the
ring.


\placefigure{mom0}


In Fig.\,\ref{hirad} we plot the azimuthally--averaged column density
versus radius, taking the center of the ring as the origin (i.e., the
position of lowest \ion{H}{1} column density as indicated by a cross
in Fig.\,\ref{mom0}). As we will explain below, this does {\it not}
coincide with the dynamical center of the galaxy. Figure \ref{hirad}
shows that the ring peaks at a radius of 52$\arcsec$, corresponding to
a diameter of $\sim$ 1.7\,kpc, followed by a fairly steep decline.
From a radius of 125{\arcsec} onwards the \ion{H}{1} profile levels
off. A determination of the \ion{H}{1} mass located in the ring by
integrating the radial profile down to a radius of 125{\arcsec} yields
$\sim 8\times 10^{7}\,{\cal M}_\sun$, which is nearly 75\% of the
total \ion{H}{1} mass in the system (see also Sect.\,\ref{blowout}).
The contrast in column density between the central \ion{H}{1}
depression at $\alpha_{2000} = 9^h 40^m 30^s$, $\delta_{2000} =
71{\degr} 11{\arcmin} 1\farcs8$ and the ring--like structure is about
a factor of 20, with a central, average value of $6 \times
10^{19}$\,cm$^{-2}$. In the following, we refer to this location of
lowest column density within the ring as the morphological center and
use it as the center of the ring itself. The error of this position is
only given by the beam of the \ion{H}{1} observation. However, the
ring seems to be slightly elongated in the north--south direction.
Judging from the rim alone, the genuine center of the ring--like
structure might differ from the point of lowest column density by
$\sim 10\arcsec$.


\placefigure{hirad}


\subsubsection{\ion{H}{1} velocity field}
\label{velfield}

Fig.\,\ref{mom1} shows the intensity--weighted velocity map of
{\holmi}, overlaid as contours on the integrated \ion{H}{1} map. The
projected velocity contours shown span a range between 130\,{\kms} and
150\,{\kms}, implying that {\holmi} is rotating. However, the
irregularity of the isovelocity contours shows that turbulent motions
dominate. This is confirmed by inspection of position--velocity (pV)
cuts along the kinematical major and minor axis and through the center
of the hole (Fig.\,\ref{pv}). The ring, described in the previous
section, is not seen to be expanding. However, this might be due to
the rather face--on view of {\holmi}.


\placefigure{mom1}
\placefigure{pv}


We performed a tilted--ring analysis using the \ion{H}{1} data to
derive the dynamics of {\holmi}. For this purpose, we smoothed the
\ion{H}{1} datacube to 20${\arcsec}$ resolution and used the task {\sc
 rotcur} \citep{beg89} in GIPSY. Model velocity fields were
subsequently developed with the task {\sc velfi} and subtracted from
the observed velocity field. This approach helps to set restrictions
to the parameters determined by {\sc rotcur}. The kinematic center as
well as the systemic velocity for {\holmi} were found to be well
constrained. The dynamical center of {\holmi} is located some
45${\arcsec}$ ($\sim 0.75$\,kpc) to the north of the morphological
center, at a systemic velocity of ($141.5\pm 1$)\,{\kms}. The
dynamical center is indicated by the intersection of the major and
minor axis in Fig.\,\ref{mom0} at $\alpha_{2000} = 09^h 40^m 31\fs6$,
$\delta_{2000} = 71{\degr} 11{\arcmin} 45{\arcsec}$. The uncertainty
in determining the center is of order 5{\arcsec}.

We were unfortunately not able to get stable results for the
inclination (as is usually the case for nearly face--on galaxies). Not
even the construction of model galaxies with the task {\sc galmod}
covering the inclination--scaleheight parameter space helped to
constrain its value. The position angle was determined, within the
errors, to have a constant value of 50${\degr}$, measured from north
through east. Rather than to plot the intrinsic rotational velocities
we present in Fig.\,\ref{rotcur} a plot of $V \times \sin (i)$ as a
function of radius. The rotation curve reaches a maximum near the peak
of the \ion{H}{1} ring beyond which it starts to decline. The rotation
curve flattens out again for $R \gtrsim 1.5$\,kpc, just where the \ion{H}{1}
ring ends. It should be borne in mind, though, that this might be an
artifact due to the fact that both the position angle and inclination
were kept fixed. Our results are in agreement with the earlier, low
resolution data published by \citet{tul78}.


\placefigure{rotcur}


We used Fig.\,\ref{rotcur} to set an upper limit to the inclination of
{\holmi} to $i\lesssim 14{\degr}$, by simply equating the dynamical
mass to the sum of the gaseous and stellar mass, the latter one based
on the optical blue luminosity (Sect.\,\ref{optprop}) and assuming an
${\cal M}_{stars}/L_{\rm B} = 1.0$. This is, of course, a lower limit
for the total mass (not accounting for molecular gas and dark matter;
see also Sect.\,\ref{region1sec}).

The \ion{H}{1} velocity dispersion shown in Fig.\,\ref{mom2} indicates
that there are two distinct regions in {\holmi}. In the southern part
we find values of $\sim 9$\,{\kms}, similar to the dispersion
encountered in other quiet, gas--rich (dwarf) galaxies \citep[see
e.g.,][]{sti99}. In the north--western part, however, significantly
higher values of $\sim 12$\,{\kms} are found. A visual inspection of
various spectra obtained at different positions reveals that the high
velocity dispersion is an intrinsic property of the \ion{H}{1} in
{\holmi}, and not due to multiple velocity components blending along
the line of sight. Table\,\ref{generaltable} summarizes the \ion{H}{1}
and optical characteristics of {\holmi}.


\placefigure{mom2}


\placetable{generaltable}


\section{Discussion}
\label{discuss}


\subsection{The Light Distribution}
\label{light}
In Fig.\,\ref{hi+opt} we present the Johnson--Cousins R$_c$ band and
the {\Ha} image, the latter one with superposed contours of the
integrated \ion{H}{1} emission. The extent of the \ion{H}{1} is
comparable to that of the underlying red low--surface brightness
stellar component. The optical surface brightness distribution of the
brighter component is asymmetric, being higher towards the southern
half, more or less coinciding with the \ion{H}{1} depletion. It is the
same region which shows the bluest color and hence suggests a
relatively young population (cf. Fig.\,\ref{holmi_rgb}). At a level of
$\mu_{\rm B}=25$\,mag/\sq{\arcsec} (D$_{25}$) this population extends
over $220\arcsec \times 60\arcsec$, or 3.7\,kpc $\times$ 1\,kpc, with
a position angle of $\sim 112{\degr}$.


\placefigure{hi+opt}


Fig.\,\ref{profiles} shows the azimuthally--averaged UBV(RI)$_c$
surface brightness profiles of {\holmi} centered on the morphological
center of the \ion{H}{1} hole. As an interesting feature we note that
the U, B, V, and R$_c$ band profiles show a change in slope at about
the D$_{25}$ diameter, with a steep exponential profile at the outer
parts of {\holmi}, and a shallower exponential disk towards the
center. The difference in slope between the inner and outer parts
becomes less pronounced with increasing wavelength. An elliptical
averaging, following the optical isophotes, leads to a similar
distribution with the kinks at the same radii.

If it were not for the absence of a bulge component, the surface
brightness profile would be classified as a Type II profile
\citep{fre70}. The kink is located at the outer edge of the
\ion{H}{1} ring, where the \ion{H}{1} column density falls below the empirical
SF threshold of $10^{21}$\,cm$^{-2}$ \citep{ski87}, as indicated by
the right--most dotted vertical line in Fig.\,\ref{profiles}. We
disregard absorption as a possible mechanism for the change in
slope. Given the low (8\% solar) metallicity, dust is expected to play
a modest role, if any. Besides, the effect becomes more pronounced as
one moves outward which would require an increasing dust opacity with
increasing radius, which is unlikely.

A more acceptable explanation is that SF has been occurring fairly
uniformly within the inner $74\arcsec$ radius. Beyond this radius,
current SF is suppressed (column densities are below the canonical
star formation threshold). As SF has occurred recently, supposedly
superimposed on an older underlying component, the light within the SF
radius is still predominantly blue, hence the rather flat and blue
color distribution. A similar trend has been observed in
M\,81\,dwarf\,A and might therefore be a general feature of low--mass
dIrr galaxies (Ott et al., in preparation).

Fig.\,\ref{b-r+halpha} compares the azimuthally--averaged B$-$R$_c$
color with that of the {\Ha} emission. Since the \ion{H}{2} regions
are distributed in a few patches, one should not take the averaging
too literally. But it does seem significant that the {\Ha} emission is
found towards the outer edge of the \ion{H}{1} ring. The current star
formation rate (SFR) is low: \citet{mil94} derive a total {\Ha}
luminosity of $4.27\times10^{38}$\,erg~s$^{-1}$ for {\holmi}, which
corresponds to a SFR of only 0.004\,${\cal M}_\sun$\,yr$^{-1}$. 

The fact that star formation is almost absent can be understood in the
framework of a universal minimum surface density threshold in
irregular galaxies \citep{gal84,ski87,tay94,zee98,hun01}. Its
value hovers typically around an \ion{H}{1} column density of
$10^{21}$\,cm$^{-2}$ which corresponds to a mass surface density of
8\,${\cal M}_\sun$\,pc$^{-2}$. There is a debate going on about its
exact value and its dependence on Hubble type and metallicity
\citep[e.g., ][]{fra86}. For larger systems, \citet{ken98} finds that
a sample of 61 normal disk galaxies show {\Ha} emission above a
surface density of typically $\Sigma_{\rm HI+H\alpha} =
10^{0.5}$\,${\cal M_\sun}$pc$^{-2}$. In other words, the threshold for
disk galaxies corresponds to $\sim 4 \times 10^{20}$\,cm$^{-2}$.  
Inspecting Fig.\,\ref{hi+opt}, SF in {\holmi} occurs where the
\ion{H}{1} column density exceeds $\sim9 \times 10^{20}$\,cm$^{-2}$ or
7\,${\cal 
 M}_\sun$\,pc$^{-2}$. Hence, {\holmi} manages to reach values close
to the empirical threshold value for dwarf galaxies in a few areas only,
i.e., coinciding with the densest regions in the ring (note that in
the case of dwarf galaxies one usually ignores the contributions by He
or H$_2$ to the mass surface density).

An alternative explanation for a star formation threshold is that of
gravitational instability. In gas--rich, spiral galaxies a thin gas
disk will undergo gravitational collapse if its gas surface density
exceeds a threshold surface density. This critical surface density is:
$\Sigma_{crit} = \alpha \frac{\kappa \sigma}{3.36 G}$ \citep{too64,
 ken89} where $\alpha$ is a constant of order unity, $\sigma$ the
1--dimensional gas dispersion, and $G$ is the gravitational constant;
$\kappa$, the epicyclic frequency, is defined as: $\kappa = 1.41
\frac{V}{R} (1 + \frac{R}{V} \frac{dR}{dV} )^{1/2}$. If we consider
only the \ion{H}{1} gas parameters (cf. Tab.\,\ref{generaltable}) and
use the radii R and velocities V given by the rotation curve and take
into account the derived inclination range at the location of the
supergiant \ion{H}{1} ring (cf. Sect.\,\ref{velfield}), we find
$20\,{\cal M_\sun}$pc$^{-2} \lesssim \Sigma_{crit} \lesssim 70\,{\cal
 M_\sun}$pc$^{-2}$. The observed mass surface density never reaches
more than $7\,{\cal M_\sun}$pc$^{-2}$, i.e., well below what would be
needed for SF to take place.

There are good reasons to argue that a thin disk approximation is not
valid in the case of thick dwarf galaxies and alternative approaches
are required \citep{hun96,hun98,wil98,hun01}. However, observed values
for $\Sigma_{gas}$, even when corrected for He, lie comfortably below
the threshold indicating that the gas disk is stable against
gravitational instability and as a result does not experience large
scale star formation. Both formalisms, a universal threshold and a
gravitational instability criterion, seem to be able to explain the
lack of current star formation in {\holmi}; based on the current data
we find that it is not possible to decide which of the mechanisms is
the dominant one.


\placefigure{profiles}
\placefigure{b-r+halpha}

Most of the {\Ha} emission in {\holmi} lies on the outer rim of the
large superbubble. This is in contrast to what has been found to be
the case for the shells of, e.g., Holmberg\,II \citep{puc92} and
IC\,2574 \citep{wal99}, where some of the brightest
\ion{H}{2} regions tend to coincide with the inner rims of the
shells. We'd like to note that there is an obvious difference,
however. In Holmberg\,II and IC\,2574 most of the rings showing {\Ha}
emission are still expanding whereas the shell in {\holmi} has
virtually stalled. The faint {\Ha} seen just to the outside of the
rim of {\holmi}'s huge \ion{H}{1} shell might thus be due to a
sufficient amount of material being piled up at the front of the
shell.


\subsection{Color--Magnitude Diagrams}
\label{cmdchap}

The seeing in our CCD images is $\sim 1\farcs4$ which corresponds to a
linear resolution of 24\,pc at the distance of {\holmi}. This means
that only the largest open clusters are likely to be resolved. As a
consequence, we are not able to distinguish between most of the
small stellar clusters and individual stars.

To study exclusively objects belonging to {\holmi} it is necessary to
discriminate them from unrelated foreground/background sources
(Galactic stars/unresolved galaxies). This was done assuming that all
sources outside a radius of $\sim$ 1\farcm9, centered on the
\ion{H}{1} depression (cf. Sect.\,\ref{light}), are unrelated objects.
The vast majority of these objects have colors (U$-$B)$\gtrsim$0.5,
(B$-$V)$\gtrsim$0.8, and (V$-$R)$\gtrsim$0.6. We assume that objects
with the same colors that fall within the aforementioned radius of
$\sim$ 1\farcm9 are background or foreground sources, too, and we
exclude them from further analysis. The color--magnitude diagrams of
the remaining point sources which likely belong to {\holmi} are
shown in Fig.\,\ref{cluster} and \ref{cmds}.

First, we start with assuming that all observed objects within
{\holmi} are unresolved stellar clusters and compare their properties
with the well--studied cluster population in the Large Magellanic
Cloud (LMC), using the compilation of LMC clusters by \citet{bic96}.
In Fig.\,\ref{cluster} we overlay their sample on the (U$-$B, B$-$V)
color--color diagram and the (V, B$-$V) color--magnitude diagram (CMD)
of our objects. To make a proper comparison, we adjusted the reddening
of the LMC cluster compilation to the reddening towards {\holmi} and
scaled the brightness according to the distance of {\holmi}. Adding
up the point sources above a V magnitude of $22.2+1.5\,$(B$-$V) ---
indicated by a solid line in Fig.\,\ref{cluster} --- yields a number
of 40 for {\holmi} and 593 for the LMC. In other words, about 15 times
more clusters would have been observed if the LMC were placed at the
distance of {\holmi}. Even correcting for the fact that the LMC is a
factor of 3 more gas rich than {\holmi} \citep[LMC \ion{H}{1} mass:
$3.1\times 10^{8}\,{\cal M}_\sun$,][]{luk92}, and scaling the number of
stellar clusters with this factor, leads still to 5 times more
clusters in the LMC, compared with the number of bright point sources
in {\holmi}. The CMD of {\holmi} most likely consists of both stellar
clusters and single stars. Therefore the cluster abundance of
{\holmi} might be even lower. This is supported by the color--color
diagram in Fig.\,\ref{cluster}, where in general the LMC clusters with
(B$-$V)$\approx 0.6$ are fainter in the U band compared to the point
sources of {\holmi}.

A reason for the difference might be the ongoing interaction of the
LMC with the Small Magellanic Cloud and the Milky Way. \citet{ken87}
and \citet{bus87}, for example, find an enhanced SFR in galaxies with
a close companion. {\holmi}'s neighbors, the M\,81 triplet galaxies
(cf. Sect.\,\ref{sec:rampressure}), are at a much larger projected
distance for which reason the effect of tidal interactions is thought
to play a less prominent role. Based on our data we can only say
something about the bright and presumably rather young stellar
clusters. The number of old clusters in the LMC and in {\holmi} might
be the same, after adjusting for the sizes of the two galaxies.


\placefigure{cluster}


Second, we discuss the consequences of the assumption that all point
sources are stars. We overlaid the isochrones by \citet{ber94} on the
CMDs in Fig.\,\ref{cmds}. The chosen metallicity of Z=0.001 is close
to the spectroscopically derived 12+log(O/H)=7.7 abundance of the
brightest \ion{H}{2} region \citep{mil96}. Note, that one cannot
directly compare the metallicity given by the oxygen abundance with
the heavy element abundance described by Z. For the Magellanic Clouds,
however, \citet{gil91} find a nearly solar [O/Fe]$\sim -0.3$. The
isochrones have been reddened according to Galactic foreground
extinction \citepalias{sch98} in order to match the uncorrected
distribution of stars in the CMDs. There is some evidence of the main
sequence (MS) in all CMDs, although it is not the most populous region
in the diagrams. \citet{apa95} found that crowding effects generally
shift blue stars to the red and red stars to the blue. This finding
suggests that the MS might be more populated than it appears to be in
the CMDs. Together with the existence of \ion{H}{2} regions, this is
an indicator for young, not very evolved stars. However, the bulk of
stars is at least older than 30\,Myr ($\log\,t=7.5$). In addition, we
do see stars in the helium--burning (HeB) blue--loop phase. Due to the
described crowding effects, the number of HeB stars might be
overestimated.


\placefigure{cmds}


The filled circles in the (V, U$-$B) diagram indicate all stars within
a radius of 0\farcm9, well \emph{inside} the main \ion{H}{1} ring
feature. These stars appear to follow isochrones between 15 and
30\,Myr ($\log\,t=7.2$--$7.5$), which is about half of the age derived
from the \ion{H}{1} properties of the ring (see also
Sect.\,\ref{creation}). All these objects are located at a mean
radius of $\sim 660$\,pc from the morphological center, following the
ring--shaped \ion{H}{1} morphology (cf. boxes in Fig.\,\ref{histars}).
The small time window supports the interpretation for the surface
brightness profiles given in Sect.\,\ref{light}. If the supergiant
\ion{H}{1} shell has indeed been created by SF, part of the triggered
population is most likely this population of stars.

In Fig.\,\ref{histars} the location of the youngest stars is shown.
These stars were selected to have (U$-$B)$\lesssim -0.7$,
(B$-$V)$\lesssim -0.1$, and V$\lesssim 23$. They neither lie at the
\ion{H}{1} depression nor at the peak of the \ion{H}{1} ring, but
rather they are associated with the currently visible \ion{H}{1} ring.
This we interpret as more evidence for the \ion{H}{1} ring being the
likely remnant of a superbubble which by its expansion has been the
trigger for subsequent SF and therefore has defined the recent SF
history of {\holmi}.


\placefigure{histars}


In contrast, older and redder stars do not show any preferred location
in {\holmi}. \citet{dyk98} found a similar behavior in the nearby
dIrr Sextans\,A, which shows a comparable \ion{H}{1} morphology. In
Sextans\,A the HeB blue--loop stars are found mainly inside the
\ion{H}{1} ring--structure. Comparing the stellar photometry of
{\holmi} with Sextans\,A is difficult, because {\holmi} is at about
twice the distance. The resulting severe crowding effects can only be
alleviated, e.g., by the use of space--based telescopes or adaptive
optics.


\subsection{The Thickness of {\holmi}}
\label{thickness}

In order to set limits on the thickness of the \ion{H}{1} layer of
{\holmi} we proceed as follows. We assume that the \ion{H}{1}
distribution perpendicular to the disk is Gaussian--shaped; for a
face--on galaxy this yields

\begin{equation} 
N_{{\rm HI}} = \sqrt{2 \pi}\,h\,n_{{\rm HI}},
\label{ncd}
\end{equation}

\noindent where $N_{{\rm HI}}$ is the \ion{H}{1} column density, $h$ the
1$\sigma$ scaleheight and $n_{{\rm HI}}$ the \ion{H}{1} particle
volume density at the midplane of the disk. The equation connecting
the scaleheight to the velocity dispersion is given by \citet{kru81}:

\begin{equation}
h(R) = \frac{\sigma_{gas}}{\sqrt{4\,\pi\,G\,\rho_{tot}(0,R)}}.
\label{kelformula}
\end{equation}

\noindent Here, $\sigma_{gas}$ is the one--dimensional \ion{H}{1} velocity
dispersion, $G$ the universal gravitational constant, and
$\rho_{tot}(0,R)$ the total mass density in the disk at radius $R$ and
$z=0$. Using both equations we derive

\begin{eqnarray}
h&=&(\sqrt{8 \pi} G m_p)^{-1}\, \frac{\sigma_{gas}^2}{\frac{\rho_{tot}}
{\rho_{{\rm HI}}} N_{{\rm HI}}} \notag \\
\,\,&=&\,5.79\times10^{21}\,
\left(\frac{\sigma_{gas}}{\rm km\,s^{-1}}\right)^{2}\,
\left(\frac{N_{\rm HI}}{\rm cm^{-2}}\right)^{-1}\,
\left(\frac{\rho_{\rm HI}}{\rho_{tot}}\right)\,{\rm pc},
\label{height} 
\end{eqnarray}

\noindent with $m_p$ the proton mass. For an average \ion{H}{1} 
column density of $3.9 \times 10^{20}$\,cm$^{-2}$, an average velocity
dispersion of 9\,{\kms}, and adopting for ${\rho_{tot}}/{\rho_{{\rm
   HI}}}$ a constant ratio of 2.2 (assuming
${\rho_{tot}}/{\rho_{{\rm HI}}}={\cal M}_{tot}/{\cal M}_{\rm HI}$;
${\cal M}_{tot}={\cal M}_{\rm HI}+{\cal M}_{\rm He}+{\cal M}_{\rm
 stars}$, with ${\cal M}_{stars}/L_{\rm B} =1$, a correction for the
contribution of helium of ${\cal M}_{\rm He}=0.3{\cal M}_{\rm HI}$ and
ignoring any dark matter contribution), we derive an upper limit of $h
\lesssim 550$\,pc, corresponding to a FWHM thickness of $\lesssim
1300$\,pc.\\ This is comparable to what has been found by \citet{puc92},
\citet{wal99}, and \citet{wal01} for the dIrrs Holmberg\,II, IC\,2574,
and DDO\,47. From this and Eq.\,\ref{ncd} we derive an average
\ion{H}{1} volume density of $n_{{\rm HI}}\gtrsim 0.10$\,cm$^{-3}$ at
the midplane of the disk.

\subsection{Evolution of the supergiant \ion{H}{1} shell}
\label{creation}


\subsubsection{Energy needed to create the shell}
\label{energysec}

As mentioned in Sect.\,\ref{intro}, the standard scenario to explain
shell structures is the deposition of vast amounts of energy via SN
explosions of the most massive stars within a region of star
formation, such as an OB association or a stellar cluster. {\holmi}
shows evidence for this scenario as we see a concentration of bright,
blue stars within the \ion{H}{1} ring (see Sect.\,\ref{optprop}). The
effect of the energy released by one SN in the post--Sedov phase on
the ambient medium has been evaluated by \citet{che74}. The kinetic
energy deposited in the ISM can be derived from

\begin{equation}
E = 5.3 \times 10^{43} \left(\frac{n_0}{{\rm cm}^{-3}}\right)^{1.12}\, 
\left(\frac{R}{{\rm pc}}\right)^{3.12}\, 
\left(\frac{v}{{\rm kms}^{-1}}\right)^{1.40}\, {\rm erg},
\label{chevalier}
\end{equation} 

\noindent where $n_0$ is the volume density of the gas in the ambient medium,
$R$ the radius, and $v$ the expansion velocity of the bubble.

It has become standard practice to take this equation and apply it to
the input of an entire stellar cluster. Since we do not see the ring
expanding, we can assume that the shell has more or less stalled at
the radius where the expansion velocity became comparable to the
\ion{H}{1} dispersion of the ambient medium (9\,{\kms}). However, due
to the rather low inclination of {\holmi}, we cannot exclude some
expansion of the supergiant shell which is not visible in the data.
As volume density we take the previously calculated value for the
average \ion{H}{1} volume density of 0.10\,cm$^{-3}$
(Sect.\,\ref{thickness}). Based on Fig.\,\ref{hirad} (see also
Sect.\,\ref{hidistr}) we derive a radius of the shell of 850\,pc. This
leads to an energy requirement of $\lesssim 1.2 \times 10^{53}$\,erg
to create the supergiant shell, which is equivalent to the kinetic
energy of $\sim 120$~type~II SNe (this value increases by a factor of
about two when taking into dark matter into account; see
Sect.\,\ref{region1sec}).


\subsubsection{Age of the shell}
\label{agesec}

We approximated the evolution of the shell by comparing it to models
of a Sedov expansion phase \citep{sed59,mac88,ehl97}. The radial
evolution $R(t)$ of a supergiant shell which is driven by a SN rate
$\dot{N}_{{\rm SN}}$ (each SN provides a kinetic energy $E_{{\rm
SN}}$) in a medium with the particle volume density $n_0$ and a mean
molecular mass $\mu$ is given by

\begin{equation*}
R(t)=53.1\,\left[\left(\frac{\dot{N}_{{\rm SN}}}{{\rm Myr^{-1}}}\right)
\left(\frac{E_{{\rm SN}}}{10^{51}{\rm{erg}}}\right)
\right]^{1/5}\,\cdot 
\end{equation*}
\begin{equation}
\cdot \left[\left(\frac{\mu}{1.3}\right)^{-1}
\left(\frac{n_0}{{\rm cm^{-3}}}\right)^{-1}\right]^{1/5}
\,\left(\frac{t}{{\rm Myr}}\right)^{3/5}\,{\rm pc}
\label{eq:radialevol}
\end{equation}

(note that the expansion velocity goes as $V(t)=\dot{R}(t)$).

If we assume that all of the 120 SNe (Sect.\,\ref{energysec}) went off
within $4\times10^{7}$\,Myr, the longest period for massive stars to
exist \citep[see, e.g.,][]{lei99}, we derive an
average SN rate of 3 Myr$^{-1}$. Furthermore, we will use again a
density of 0.10\,cm$^{-3}$ (Sect.\,\ref{thickness}), a mean molecular
weight of 1.3, and an average energy of 10$^{51}$\,erg per SN. If
break--out occurs, energy input into the shell becomes ineffective, and
the structure enters the ``snowplow'' phase, i.e., an expansion with
conservation of momentum. \citet{mcc87} find that for this stage of
the evolution of a supergiant shell the radius will increase as

\begin{equation}
R(t)=R_c\,\left({\frac{t}{t_c}}\right)^{1/4}
\end{equation} 

\noindent beyond a critical radius $R_c$ and a critical time $t_c$, 
which is the radius and the time of the shell right at the moment
where the shell breaks out of the disk (see also
Sect.\,\ref{blowout}). Again, the derivative of this equation yields
the velocity of this phase. Fig.\,\ref{shellsim} shows the predicted
radius and velocity at break--out for several critical variables. The
line marked ``SN driven'' represents the case without break--out. The
line shows the steady growth of the shell as a function of time. The
labels '33,' '24,' etc., represents the expansion velocity of the
shell at different ages. The family of lines marked ``snowplow''
represent the growth of the shell (at a lower rate) after it has
broken out of the disk. Note that these calculations have been
performed for a single shell with a central event. Propagating SF and
simultaneous SF within the hole as suggested by the CMDs and the
surface brightness distribution (Sect.\,\ref{cmdchap} and \ref{light})
changes the energy and the locations of energy input and therefore the
age calculations (see also Sect.\,\ref{blowout}).


\placefigure{shellsim}


The age of the supergiant shell can be estimated using various approaches:

\begin{itemize}
  
\item An upper limit for the age of this structure can be obtained by
  simply dividing the radius of the shell (0.85\,kpc) by the limit of
  the expansion velocity, 9\,{\kms}, which leads to 90\,Myr.
  
\item The dotted line in Fig.\,\ref{shellsim} shows the observed
  radius of the shell. Since we found the \ion{H}{1} 1$\sigma$
  scaleheight to be $\lesssim 550$\,pc (Sect.\,\ref{thickness}) we
  assume that the shell broke out of the disk of {\holmi}, a
  conclusion which is supported by the overall morphology (cf.
  Sect.\,\ref{blowout}). The horizontal dashed line of
  Fig.\,\ref{shellsim} indicates the size of the shell at blow--out.
  This event should have taken place at an age of $\sim 15$\,Myr. The
  intersection of the current size of the shell (dotted line) with the
  evolutionary track of the snowplow phase (thick line) leads to a
  total age estimate of $\sim 100$\,Myr.
  
\item The stars within the \ion{H}{1} shell of {\holmi} and visible in
  the U band are located at a mean radial distance of $\sim 660$\,pc
  (the open squares in Fig.\,\ref{histars}). This radius was achieved
  by the shell when it was $\sim$ 35\,Myr old (see
  Fig\,\ref{shellsim}, the thicker drawn evolutionary track). Adding
  the maximum age of these stars (30\,Myr, see Sect.\,\ref{cmdchap})
  yields an age of $\sim 65$\,Myr.
  
\item If we assume that the shock front creating the ring started as
  an infinitesimally thin shock and assume that this shock has
  broadened over time, we find that for an equivalent broadening speed
  of 9\,\kms\ it would take some 70\,Myr to expand isotropically by
  0.6\,kpc (half of the FWHM of the ring).  
  
\item Comparing the integrated light of the remnant of the central
  cluster with the models provided by \citet{lei99} predicts, based on
  a surface color of $\mu_{\rm U}(0)-\mu_{\rm B}(0)=-0.41$, and
  adopting a metallicity of Z=0.001 and a Salpeter IMF, an age of
  $\sim 60$\,Myr.

\end{itemize}

\noindent Given the uncertainties in each of the derived values, 
we conclude an age of the supergiant shell of $80 \pm 20$\,Myr.
Table\,\ref{ringtable} summarizes the properties of the \ion{H}{1}
ring.
     

\placetable{ringtable}


\subsection{Ram pressure effects}
\label{sec:rampressure}
An obvious aspect of the global optical and \ion{H}{1} morphology is
the apparent lopsidedness. We speculate that ram pressure may play a
role in causing this asymmetry, much as was proposed by \citet{bur01}
in the case of Holmberg\,II. If this is the case, {\holmi} must be
falling towards the strongly interacting M\,81 triplet consisting of
M\,81, M\,82, and NGC\,3077 \citep[for a description see][]{yun94},
which lies to the south--east at a projected distance about twice the
size of the triplet (distance {\holmi} $\leftrightarrow$ M\,82: $\sim
120$\,kpc; M\,82 $\leftrightarrow$ NGC\,3077: $\sim 80$\,kpc). Ram
pressure effects may have compressed the gas in the south--eastern
part of {\holmi}, raising it above the critical level for SF. It is
interesting to note in that context the main {\Ha} emission is
detected in the same south--eastern quadrant, i.e., the ``leading
edge'' of {\holmi}. This is somewhat similar to what \citet{deb98}
found for the LMC. On the opposite side, the north--west, one would
expect higher turbulence \citep[cf. the Coma cluster;][]{gun72}. The
linewidth distribution as shown in Fig.\,\ref{mom2} corroborates this
scenario. However, it may also be that the lopsided distribution of
\ion{H}{1} in {\holmi} is due to a (weak) tidal pull by the
same M81 triplet. We consider detailed calculations of either of the
two possibilities beyond the scope of this paper.


\subsection{The south--east of {\holmi}}
\label{region1sec}

Besides the big \ion{H}{1} hole, {\holmi} possesses several smaller
features which are visible in the channel maps (cf.
Fig.\,\ref{channmap}). The bulk of these structures are
noise--dominated, so we refrain from listing all possible cavities. A
prominent feature, however, is located in the south--eastern part of
the ring where we find the highest \ion{H}{1} column densities. The
box marked in Fig.\,\ref{mom0} is blown up in Fig.\,\ref{region1},
where we show the \ion{H}{1}, the Johnson B--band and {\Ha} emission
for comparison. This particular \ion{H}{1} shell has a diameter of
0.5\,kpc and is filled by blue stars, lending support to the picture
that massive SF (and the resulting SN explosions) is shaping the
surrounding ISM. The \ion{H}{1} column density drops by a factor of
$\sim 3$ towards the center of this ring ($N_{{\rm HI}}= 6\times
10^{20}$\,cm$^{-2}$; $\alpha_{2000} = 9^h 40^m 35^s$, $\delta_{2000}
=71{\degr} 10{\arcmin} 00{\arcsec}$). A visual inspection of pV cuts
at different position angles centered on the shell, as well as various
spectra in the natural weighted cube, reveals that the shell is still
slightly expanding at $\sim 6.5$\,{\kms}. Following
Sect.\,\ref{creation} we derive for a diameter of 0.5\,kpc and a
density of 0.10\,cm$^{-3}$ an energy requirement of $E = 1.7 \times
10^{51}$\,erg, corresponding to only about two supernovae. An upper
limit to the age of this shell is set by the life expectancy of the
least massive star still to go off as supernova, or 40\,Myr$^{-1}$. If
we use the same Sedov expansion phase model as in Sect.\,\ref{agesec},
we estimate for a SN rate of 0.05\,Myr$^{-1}$ an age of 15--20\,Myr.
In contrast to the location of the \ion{H}{2} regions on the outer rim
of the huge superbubble (cf. Sect.\,\ref{light}), we find {\Ha}
regions at the inside edge of the small shell, similar to what has
been found for other dIrrs \citep{puc92,wal99}. For a velocity of
6.5\,{\kms}, the rim of the small \ion{H}{1} shell needs some
10--15\,Myr to pass a given point in space. A comparison with the
derived age might then explain why the \ion{H}{2} regions are located
at the inner rim of this structure.

The detection of expansion in the z direction of the shell implies
that it has not broken out of the disk yet. The scaleheight of the
\ion{H}{1} layer must therefore have a lower limit of $h\gtrsim
250$\,pc, assuming an isotropic expansion. By applying
Eq.\,\ref{height}, we find an upper limit for the total mass density
of $\rho_{tot}\approx 5 \rho_{{\rm HI}}$ within {\holmi}. If visible
matter traces the dark matter content, we can estimate a total mass
for {\holmi} of $\lesssim 5.5 \times 10^{8}\,{\cal M}_{\sun}$, which
corresponds to a limit for the dark matter of $\lesssim 3.1 \times
10^{8}\,{\cal M}_{\sun}$. Given the relatively small amount involved,
this could perhaps be accounted for by molecular gas, dust, or even a
higher ${\cal M}_{stars}/L_{\rm B}$ ratio, in which case no dark
matter would be required to explain the observed rotation curve.
Adopting our curve leads to an inclination of $i\gtrsim 10\degr$.
Employing Eq.\,\ref{ncd} yields an upper limit for the particle volume
density of $\sim$0.20\,cm$^{-3}$. We can now use this value to
estimate an upper limit to the energy calculation of the supershell
(Sect.\,\ref{energysec}): $E \lesssim 2.6\times10^{53}$\,erg, which is
equal to the energy input of about 260 SNe. The determination of the
age of the shell, based on a comparison with models (cf.
Sect.\,\ref{agesec}) does not change substantially; the energy depends
nearly linearly on the volume density in Eq.\,\ref{chevalier}. If the
density doubles, the energy follows and those two factors cancel out
in Eq.\,\ref{eq:radialevol}.

For the small shell, a particle volume density of 0.20\,cm$^{-3}$ in
the disk leads to an upper limit of the energy needed for its creation
of $3.6\times 10^{51}$\,erg, which corresponds to about 4 SNe. Again,
the estimate for its age does not change much for the reasons
described above.


\placefigure{region1} 

     
\subsection{Blow--out}
\label{blowout}
Whether or not the central big \ion{H}{1} hole has suffered
``blow--out'' (i.e., the shell breaking out of the disk \emph{and}
losing metal--enriched material to the intergalactic space) highly
depends on the total mass of the galaxy. In the model of \citet{mac99}
and \citet{fer00}, objects with visible masses below $\sim
10^{9}\,{\cal M}_\sun$ suffer blow--out. The gas would then be
distributed over a bigger volume surrounding the former host. They
argue that for this to work, the velocity of the expanding shells
would have to exceed the escape velocity of the host galaxy. In the
case of {\holmi} we do not see any \ion{H}{1} along the line of sight
to the central hole down to a column density of $\sim 6 \times
10^{19}$\,cm$^{-2}$. The ``line of sight'' effect may contribute to
the difference in the measured flux between the projected rim of a
three dimensional shell and its center. This effect can, however, only
account for a change of about a factor of two in column density and is
insufficient to describe the observed \ion{H}{1} gradient from the
ring towards the center (a factor of $\sim 20$, Fig.\,\ref{hirad}).

The calculations performed in Sect.\,\ref{agesec} and displayed in
Fig.\,\ref{shellsim} suggest that the shell broke out of the disk,
since the observed radius of the ring (850\,pc) is considerably larger
than the 1$\sigma$ scaleheight of the gaseous disk ($\lesssim
550$\,pc). An alternative explanation for the shell to stall is that
its interior simply has cooled, reducing the overpressure. The
typical timescale for such a cooling, $t_{cool}$, is described by
\citet{mcc87}:

\begin{equation}
t_{cool}=4\,\zeta^{-1.5}\,(N_{*}\,E_{51})^{0.3}\,n_0^{-0.7}\,\,{\rm Myr}
\end{equation}

\noindent where $\zeta$ corresponds to the metallicity in solar units, $N_{*}$
to the number of SN explosions, $E_{51}$ to the energy per SN in units
of $10^{51}$\,erg, and $n_0$ to the ambient density of the ISM. If we
use the parameters evaluated for {\holmi} in Sect.\,\ref{thickness}
and Sect.\,\ref{energysec} ($N_{*}=120$, $n_0=0.1$\,cm$^{-3}$), assume
an energy of 10$^{51}$\,erg per SN and the \citet{mil96}
metallicity of 8\% solar, we estimate the cooling time 
$t_{cool}=3.7$\,Gyr. This is far too long to play any role in the case
of the central shell in {\holmi}.

If the radius of a shell reaches the 1$\sigma$ scaleheight of a
Gaussian shaped halo, the shock will be accelerated perpendicular to
the disk, and metal--enriched material will be ultimately lost to
intergalactic space \citep{sil01}. In this model, neither the pressure
of the intergalactic medium nor the gravity of the host galaxy is able
to retain the swept--up gas. This is exactly the ``blow--out''
scenario which was alluded in Sect.\,\ref{intro}.

For radii smaller than the scaleheight $h$, Gaussian stratification
leads to models quite similar to the ones with uniform density. At
$R=h$ the density is 0.68 times the density in the midplane. As the
radial evolution depends on the density as $n^{-1/5}$ (cf.
Eq.\,\ref{eq:radialevol}), the radii and velocities in the z direction
before break--out will be only slightly larger then the ones derived
by the Sedov model in Sect.\,\ref{agesec}. However, if one assumes an
exponentially--shaped z distribution of the gaseous component, the
acceleration and therefore blow--out does not happen until a couple of
exponential scalelengths $h_{exp}$ \citep{sil98,sil01}, which is at
about 2.4 times the Gaussian scaleheight $h$. In this case, one would
expect the shell to be still in the Sedov expansion phase with some
piled--up gas in the z--direction. However, the latter is not
supported by the observational facts.

If blow--out indeed has taken place, it is difficult to estimate how
much material was actually lost. The ring seems to dominate the galaxy
up to a radius of 125${\arcsec}$ ($\approx 2.1$\,kpc; see
Fig.\,\ref{hirad}). What may the \ion{H}{1} distribution of {\holmi}
have looked like just before the creation of the supergiant shell? We
fit different \ion{H}{1} profiles to the outer, apparently undisturbed
region of the galaxy and extrapolated them back towards the center.
This was done for a Gaussian, a linear and an exponential \ion{H}{1}
distribution (see Fig.\,\ref{interpol}). If we integrate these
profiles, we find that the \ion{H}{1} mass corresponding to the
Gaussian fit and the one based on our observations are nearly the
same, implying that the \ion{H}{1} has only been moved around. The
linear fit underestimates the observed \ion{H}{1} flux by $\sim 30$\%.
The exponential profile predicts that $\sim$50\% of the \ion{H}{1} was
lost during the creation of the supergiant shell. Although we have no
knowledge of the nature of the \ion{H}{1} distribution of \holmi\ 
before the creation of the hole, according to \citet{tay94}, many
dIrrs show an exponential \ion{H}{1} distribution. A further advantage
in adopting such a distribution is that this profile exceeds the SF
threshold in the inner parts of {\holmi} out to a radius of $\sim
1$\,kpc, which might explain why the burst of SF was able to occur
offset from the dynamical center, as observed.


\placefigure{interpol}


\section{Summary}
\label{summary}
We present high--resolution VLA \ion{H}{1} and deep optical
UBV(RI)$_c$ as well as {\Ha} observations of the dIrr galaxy {\holmi},
a member of the M\,81 group of galaxies. We find the following:\\

\begin{enumerate}
  
\item Maps of neutral hydrogen show an impressive supergiant shell
  with a diameter of 1.7\,kpc. In addition, a wealth of small--scale
  structure is seen in the \ion{H}{1} data cube. The shell covers
  about half of the \ion{H}{1} extent of {\holmi} (total \ion{H}{1}
  size: $\sim 5.8$\,kpc) and contains a substantial fraction
  ($\sim$75\%) of its total neutral gas content (total \ion{H}{1}
  mass: $1.1\times 10^8\,{\cal M}_\sun $). We derive the scaleheight
  of the \ion{H}{1} layer to be 250\,pc $\lesssim h \lesssim 550$\,pc.
  
\item A dynamical analysis shows that turbulence plays an important
  role in {\holmi}. The velocity dispersion in the north--west ($\sim
  12$\,{\kms}) is significantly higher than the overall value of $\sim
  9$\,{\kms}. A tilted ring analysis shows that the dynamical center
  is offset by 0.75\,kpc from the \ion{H}{1} morphological center.
  With the help of a small shell, which has not broken out of the disk
  yet, we estimate an upper limit of the dark matter content of
  $\lesssim 3.1 \times 10^{8}\,{\cal M}_{\sun}$. This leads to a total
  mass of $\lesssim 5.5 \times 10^{8}\,{\cal M}_{\sun}$ and an
  inclination of $10{\degr}\lesssim i \lesssim 14{\degr}$.

\item Extrapolating backwards in time what the original \ion{H}{1}
  distribution might have looked like before the ring was formed, we
  estimate a higher neutral gas mass than currently observed.
  Comparing the results with published models requires that the shell
  has broken out of the disk. Comparing the results with the
  \ion{H}{1} scaleheight and the overall \ion{H}{1} morphology leads
  to further circumstantial evidence for the occurrence of blow--out
  in {\holmi}.
        
\item With $\mu_{\rm B} (0) \sim 24.5$\,mag/$\sq \arcsec$, {\holmi}
  belongs to the class of low--surface brightness dwarf galaxies. The
  optical extent in the R$_c$ band is comparable to the \ion{H}{1}
  size. Most of the stellar light falls within the \ion{H}{1} ring.
  Azimuthally--averaged surface brightness profiles are shallow from
  the center outwards to a radius of 1.2\,kpc (where the \ion{H}{1}
  column density falls below $\sim 10^{21}$\,cm$^{-2}$) and steepen
  beyond that radius. The luminosity profile can be explained by a
  rather uniform star formation history in the center of {\holmi}.

\item The blue luminosity of $L_{\rm B} = 1.0 \times 10^8 L_{{\rm
   B}_{\sun}}$ implies an ${\cal M}_{\rm HI}$/$L_{\rm B}$ ratio of
  1.1\,${\cal M}_\sun / L_{{\rm B}_\sun}$. We estimate a total visible
  (stars plus atomic gas) mass of $2.4\times10^8 \,{\cal M}_\sun$.

\item \ion{H}{2} regions are found on the outer rim of the \ion{H}{1}
  shell. This implies that conditions in the swept--up \ion{H}{1}
  shell are suitable for star formation to commence. Some of the
  \ion{H}{2} regions on the rim are centered on a smaller \ion{H}{1}
  shell.
    
\item At the distance of {\holmi}, bright unresolved optical sources
  can be either stellar clusters or single stars. A comparison with
  LMC clusters shows that the cluster population near {\holmi} is much
  less abundant (a factor of $\lesssim 1/5$) than that of the LMC.
  From a comparison with isochrones in the (V, U$-$B) CMD we find that
  stars within the \ion{H}{1} ring are relatively young
  ($\sim$15--30\,Myr). Not surprisingly, the youngest stars are found
  to be close to the \ion{H}{2} regions on the rim of the \ion{H}{1}
  shell.
        
\item The age of the stars within the \ion{H}{1} shell as well as
  considerations on the evolution of the gaseous rim lead to an age of
  the shell of $\sim 80\pm 20$\,Myr. The kinetic energy to create the
  supergiant shell is some $1.2 \times 10^{53}$\,erg $\lesssim E
  \lesssim 2.6\times10^{53}$\,erg (equivalent to 120--260 type II SN
  explosions).

\item The south--eastern side of {\holmi} is closest to the M\,81
  triplet and shows a steep gradient in the \ion{H}{1} distribution.
  The opposite side (north--west) shows systematically higher velocity
  dispersions. {\Ha} emission is predominantly found in the
  south--east. All of these features may be indicative of ram pressure
  within the M\,81 group. Alternatively, tidal forces might be
  shaping the structure of the ISM in {\holmi} at large galactocentric
  radii.
        
\end{enumerate}

{\holmi} is a fascinating object with an impressive supergiant
\ion{H}{1} shell. Amongst dIrrs with a similar morphology, it is one
of the very few nearby, low--mass systems which can be studied in
great detail. From our observations it seems likely that the giant
\ion{H}{1} shell was created by recent star formation within the shell
(t$\sim 80\pm20$\,Myr). We can only speculate what {\holmi} will look
like in, say, 100\,Myr and what it looked like before the creation of
the \ion{H}{1} shell. It may well be that {\holmi} was a blue compact
dwarf galaxy (BCD) in the recent past, with a centrally peaked
\ion{H}{1} distribution. The active episode of star formation and
subsequent SN explosions may have blown the supergiant \ion{H}{1}
shell that we witness today. In the future the shell may recollapse
and, in the process, ignite another burst of massive star formation
(``episodic star formation''). In any case, in--depth studies of
low--mass and low--metallicity galaxies, such as {\holmi}, help us
understand how the effects of massive star formation change the
properties of the interstellar medium. They also give us some insight
in to how low--mass galaxies may have contributed to the chemical
enrichment of the intergalactic medium when the universe was young.

\acknowledgments JO would like to thank Thomas Fritz, Ulrich Mebold
and Axel Weiss for fruitful discussions. Without the funding of the
Deutsche Forschungsgemeinschaft and the associated Graduiertenkolleg
``Magellanic clouds and other dwarf galaxies'' this work would not
have been possible. The authors also thank the anonymous referee for
valuable comments which have helped to improve the presentation of
this paper. FW acknowledges NSF grant AST96--13717. EB acknowledges
support by CONACyT (grant No.\ 27606--E). BD acknowledges the award of
a Feodor Lynen grant by the ``Alexander von Humboldt--Stiftung''.
Furthermore, we thank the staff of the Calar Alto observatory for
their kind support during the observations. This research has made use
of the NASA/IPAC Extragalactic Database (NED), which is maintained by
the Jet Propulsion Laboratory, Caltech, under contract with the
National Aeronautics and Space Administration (NASA), NASA's
Astrophysical Data System Abstract Service (ADS) and NASA's SkyView.

\clearpage
\plotone{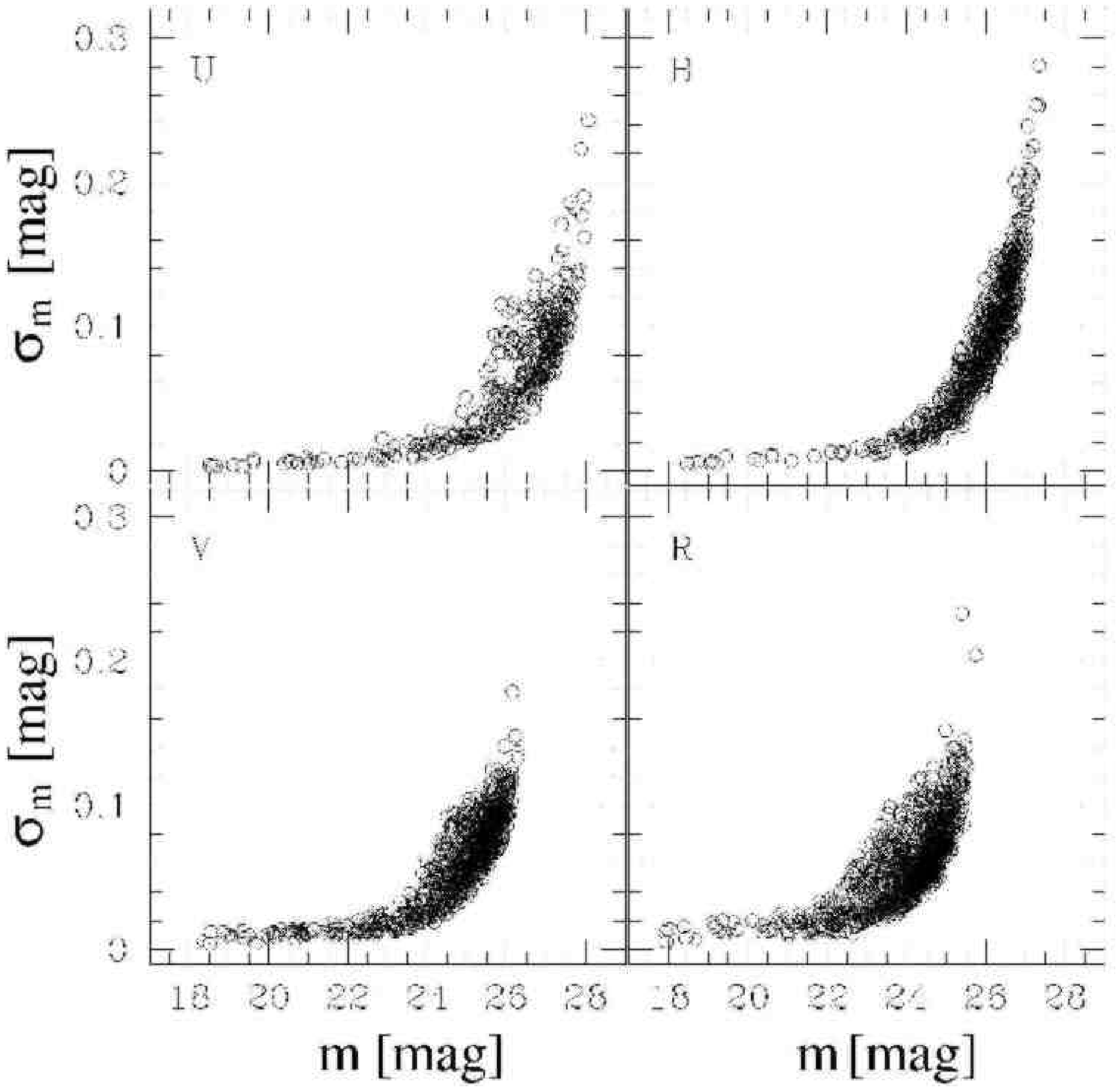}
\figcaption[ott.fig1.eps] {Formal photometric errors as a function of
  magnitude in the UBV(RI)$_c$ bands (output of {\sc allstar} in {\sc
    daophot}).  \label{formalerrors}}

\plotone{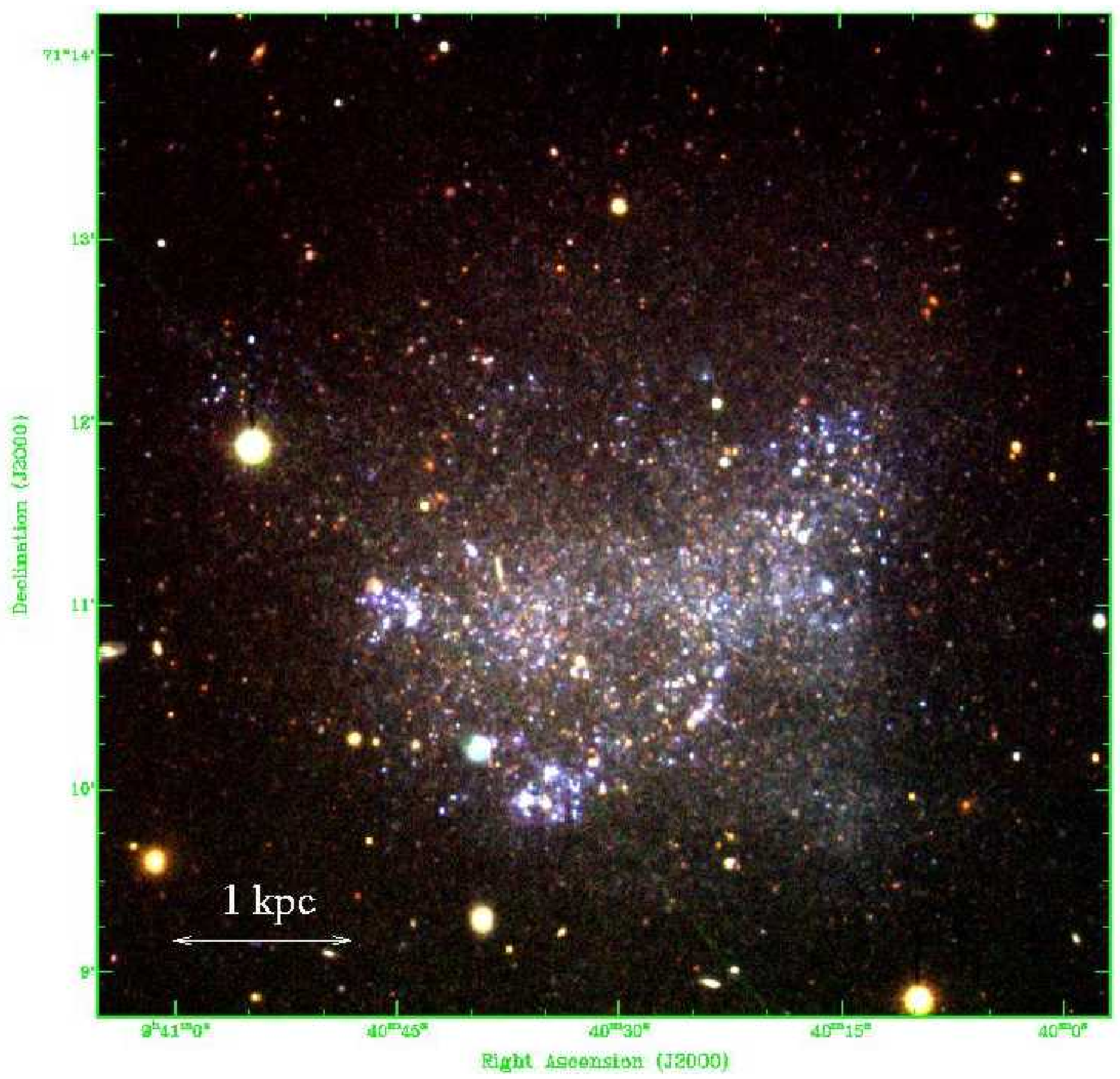}
\figcaption[ott.fig2.eps] {Three--color composite image of \holmi\ 
  based on our UBVR$_c$ CCD frames, where the blue color represents an
  average image of U and B.\label{holmi_rgb}}

\plotone{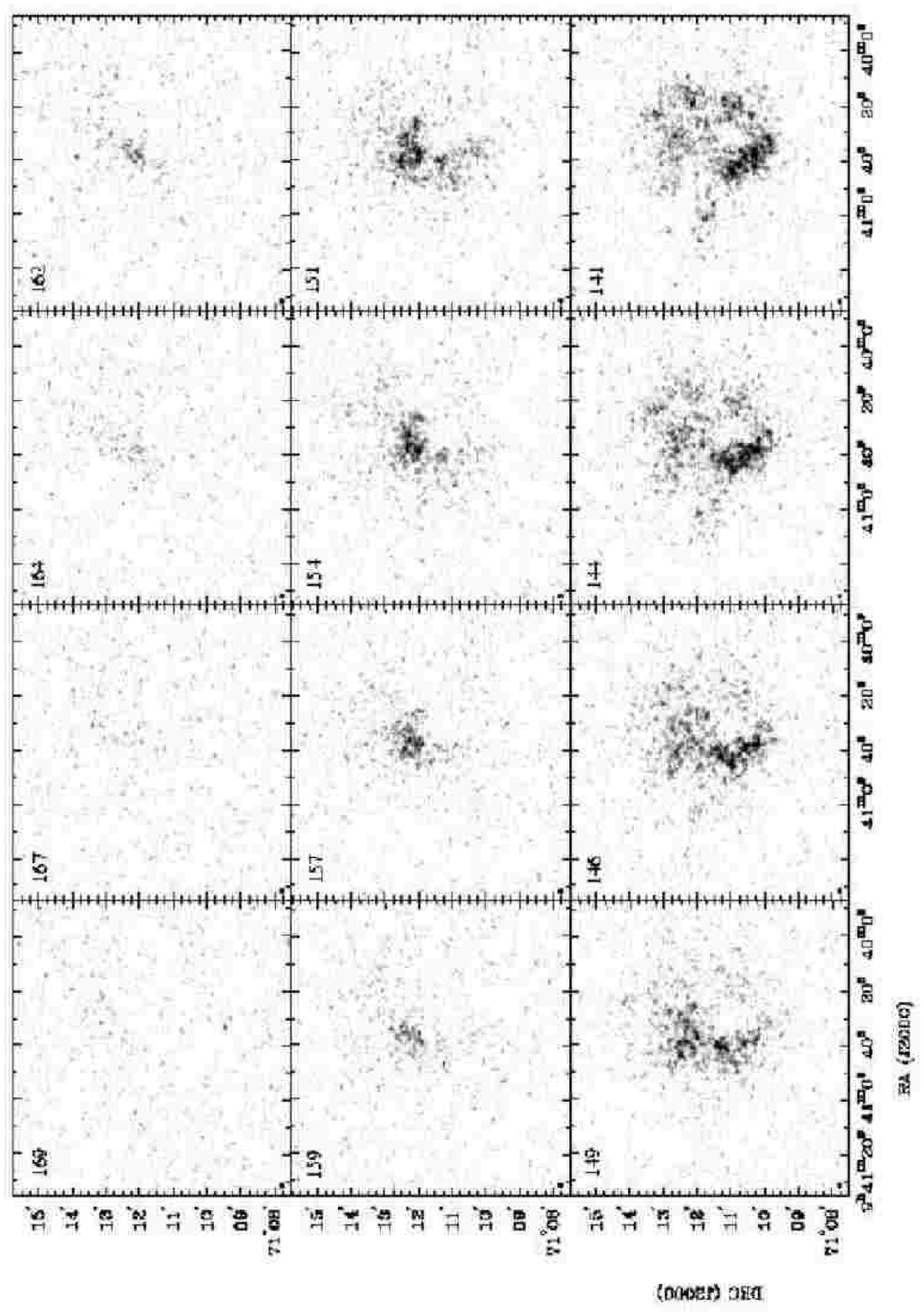}
\figcaption[ott.fig3a.eps] {\ion{H}{1} channel maps of {\holmi}. The
  shape of the beam is plotted in the lower left, and the
  (heliocentric) velocity in the upper left corner in units of {\kms}
  (1${\arcmin}$ corresponds to $\sim$1\,kpc). Note the central
  \ion{H}{1} depression in all channel maps and the fact that there is
  emission in almost every channel map in the north--west. Structures
  at smaller scales, such as rings or holes, are usually visible in
  two consecutive maps only. {\label{channmap}}}

\plotone{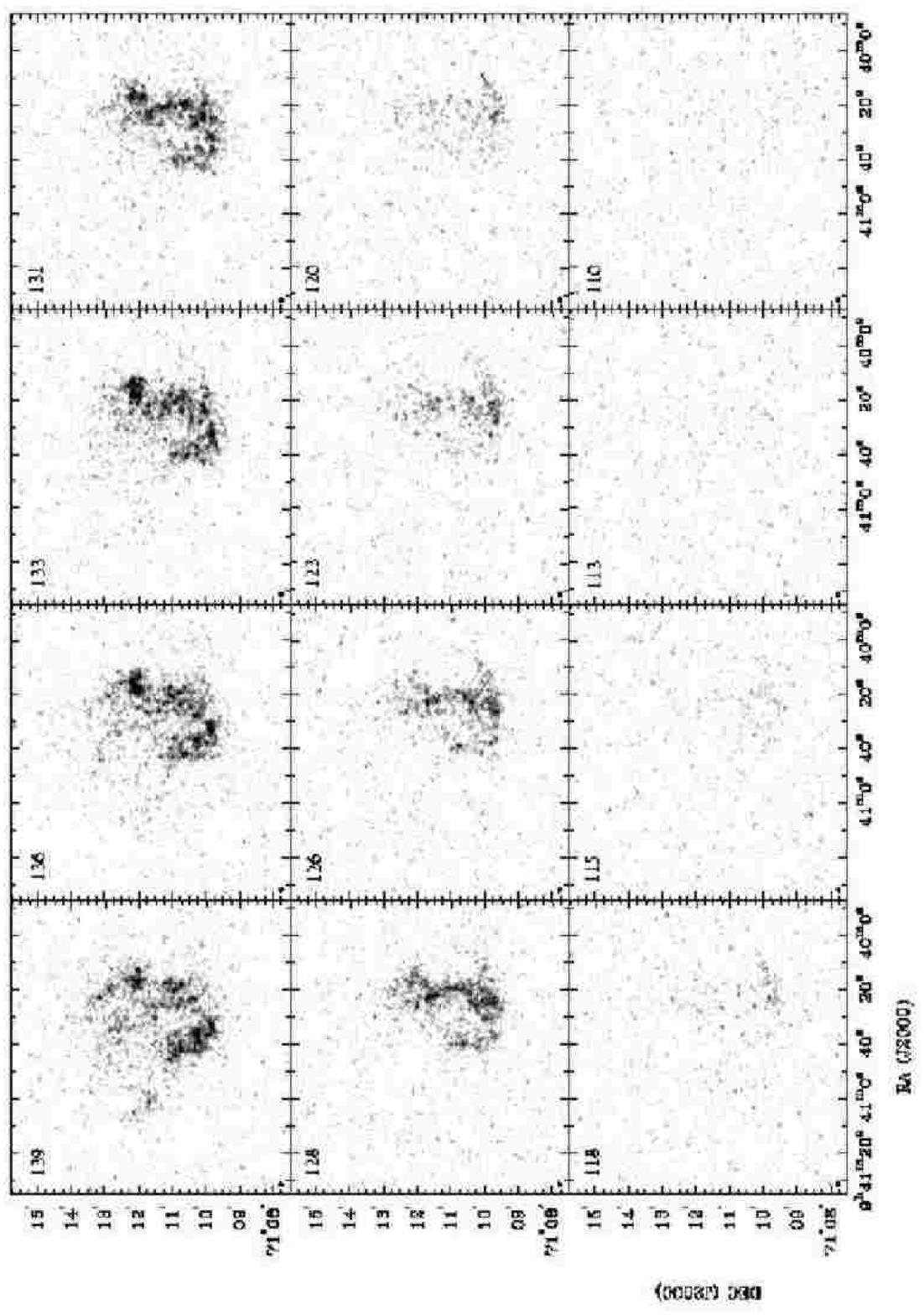}
\addtocounter{figure}{-1}
\figcaption[ott.fig3b.eps] {continued}

\plotone{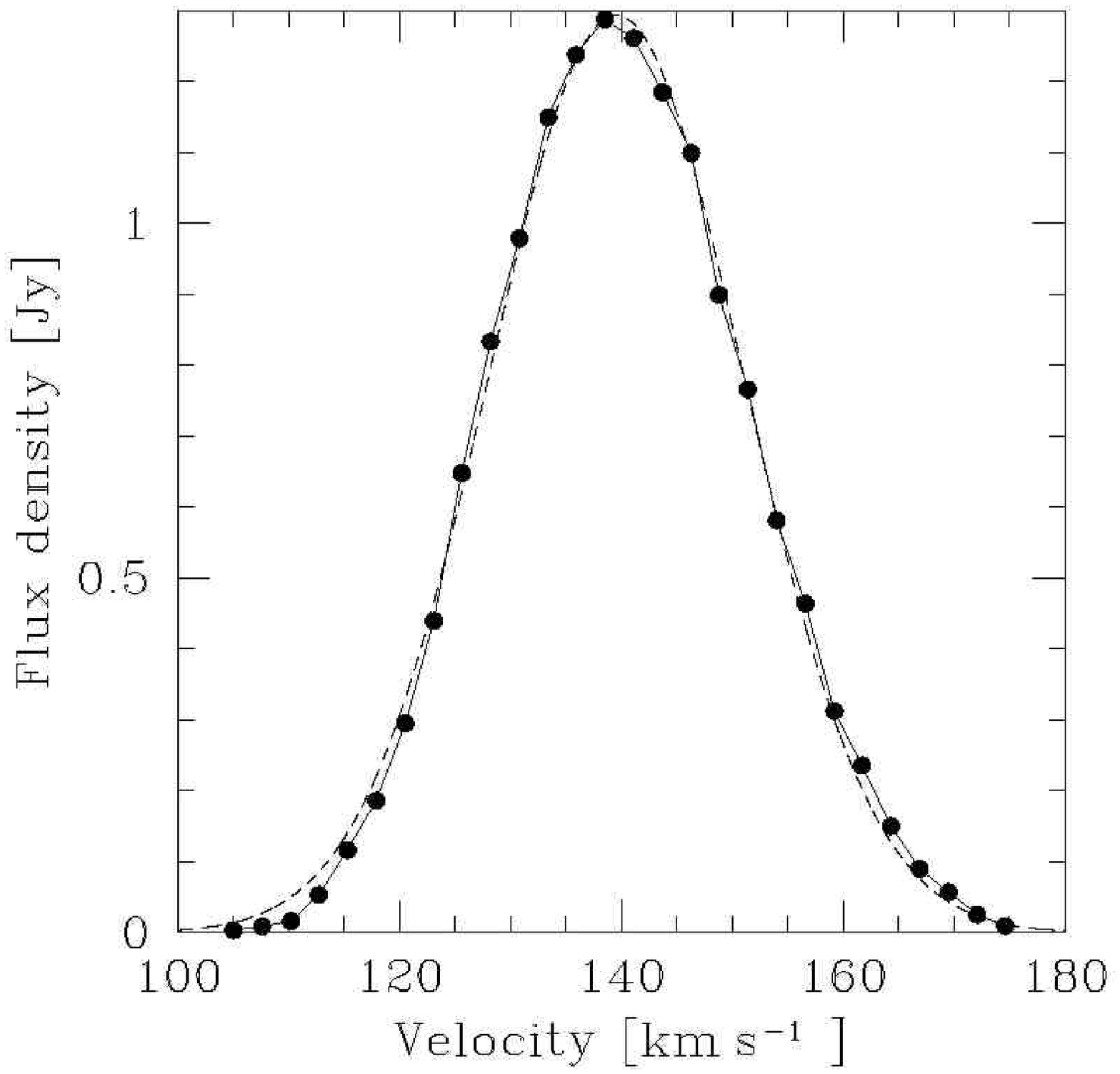}
\figcaption[ott.fig4.eps] {\ion{H}{1} spectrum of {\holmi}. The shape
  is a nearly perfect Gaussian, as illustrated by the {\it dashed
   curve}, which is a least--squares fit to the data points.
  Velocities are heliocentric.\label{spectrum}}

\plotone{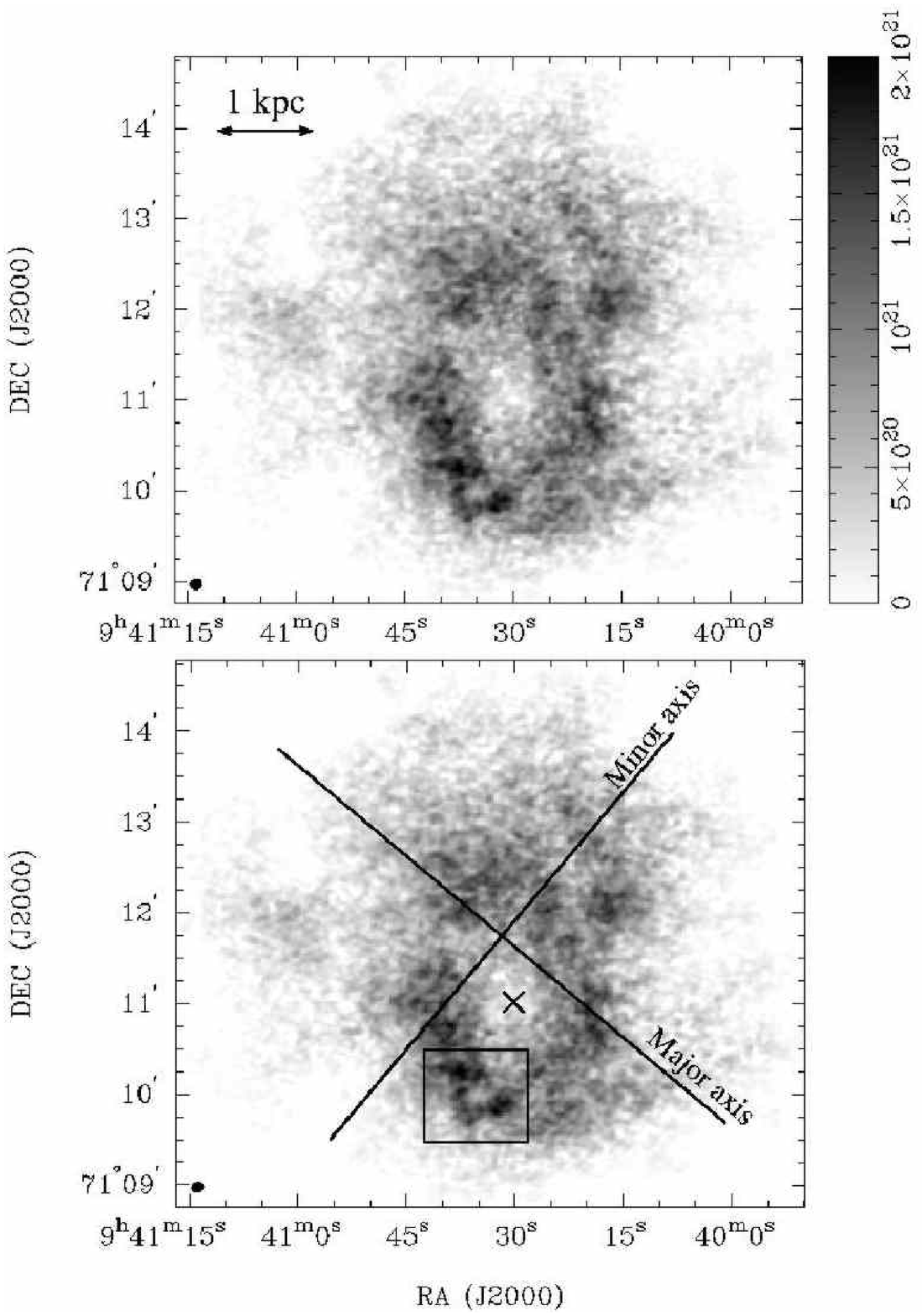}
\figcaption[ott.fig5.eps] {{\it Upper panel:} The integrated
  \ion{H}{1} emission of {\holmi}. The beam of 8\farcs2 $\times$
  7\farcs0 in size is shown in the lower left corner. {\it Lower
   panel:} The same figure as above, but with the kinematical major
  and minor axes overlaid. The cross indicates the morphological
  center, i.e., the point where the \ion{H}{1} density reaches its
  minimum. The box shows the region discussed in
  Sect.\,\ref{region1sec}. A blow--up of this area is shown in
  Fig.\,\ref{region1}.
  \label{mom0}}

\plotone{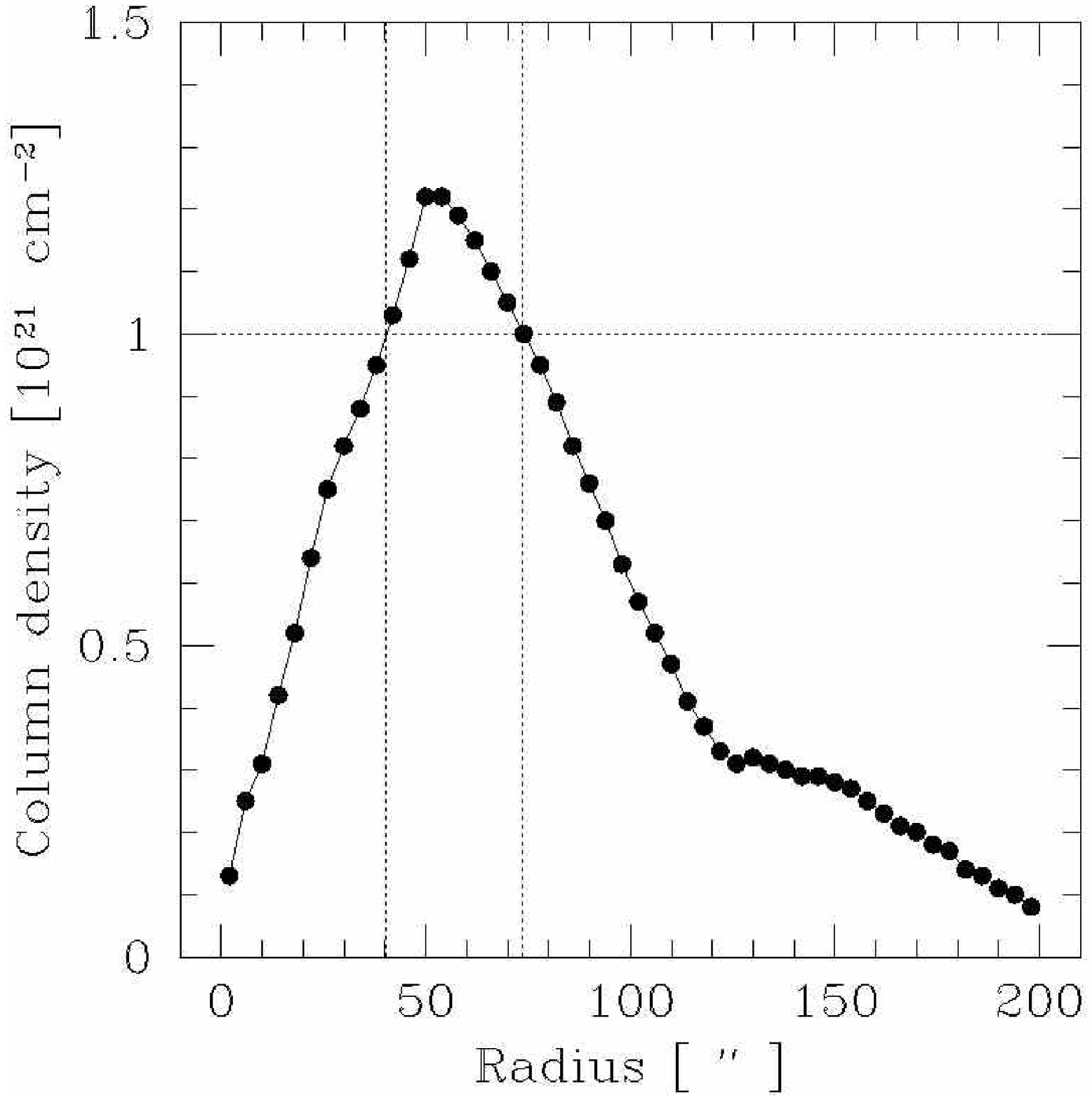}
\figcaption[ott.fig6.eps] {Azimuthally--averaged \ion{H}{1}
  distribution, taking the morphological center (cf. Fig.\,\ref{mom0})
  as the origin (1${\arcmin}$ corresponds to $\sim$1\,kpc). The
  vertical lines represent the radii where the \ion{H}{1} column
  density reaches values sufficiently high for star formation to
  commence ($N_{{\rm HI}}= 10^{21}$\,cm$^{-2}$).
\label{hirad}}

\plotone{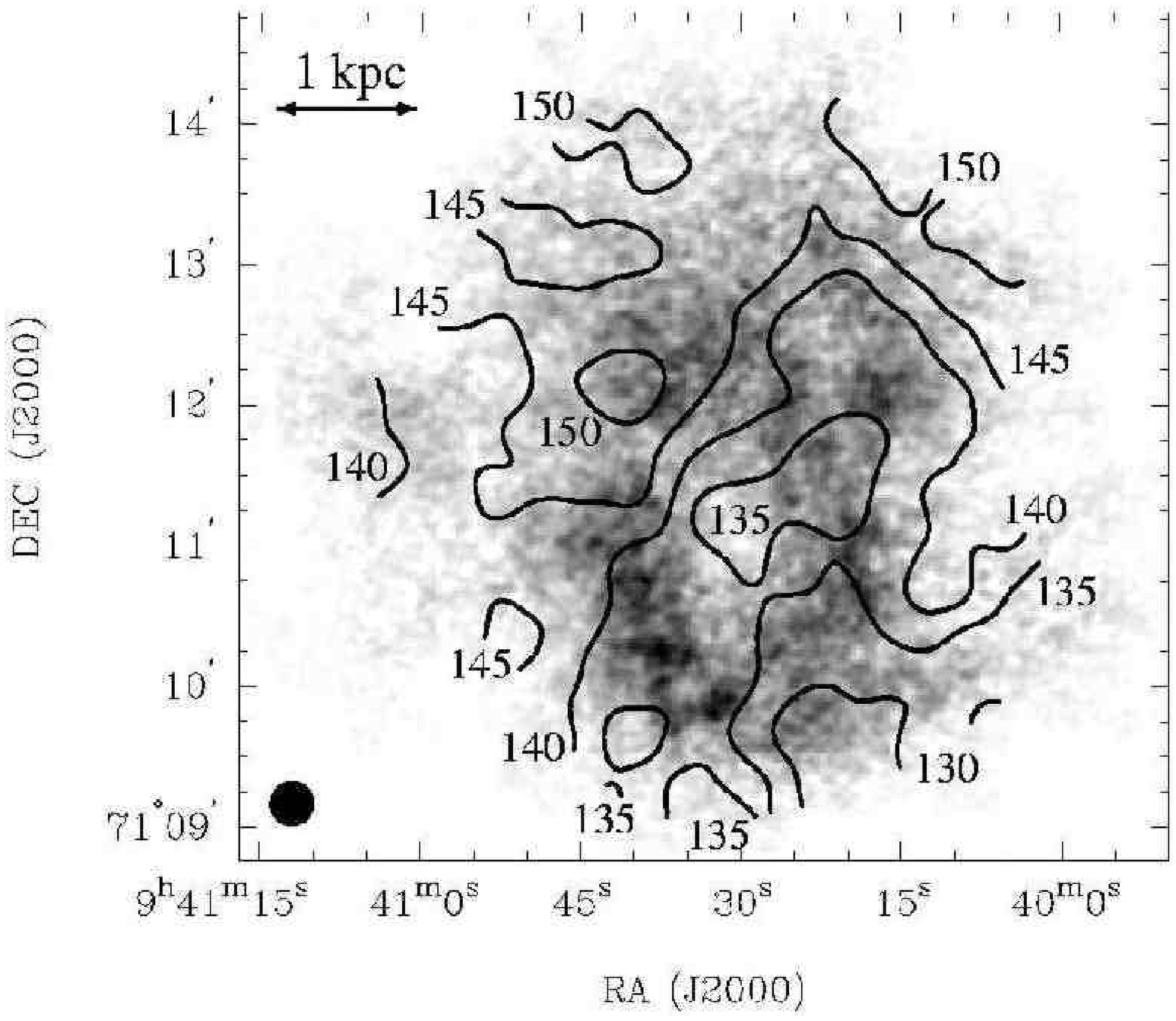}
\figcaption[ott.fig7.eps] {Velocity field of the \ion{H}{1} data of
  {\holmi}, smoothed to 20${\arcsec}$, plotted as contours overlaid on
  a greyscale representation of the integrated \ion{H}{1} map (units
  are {\kms}).\label{mom1}}

\plotone{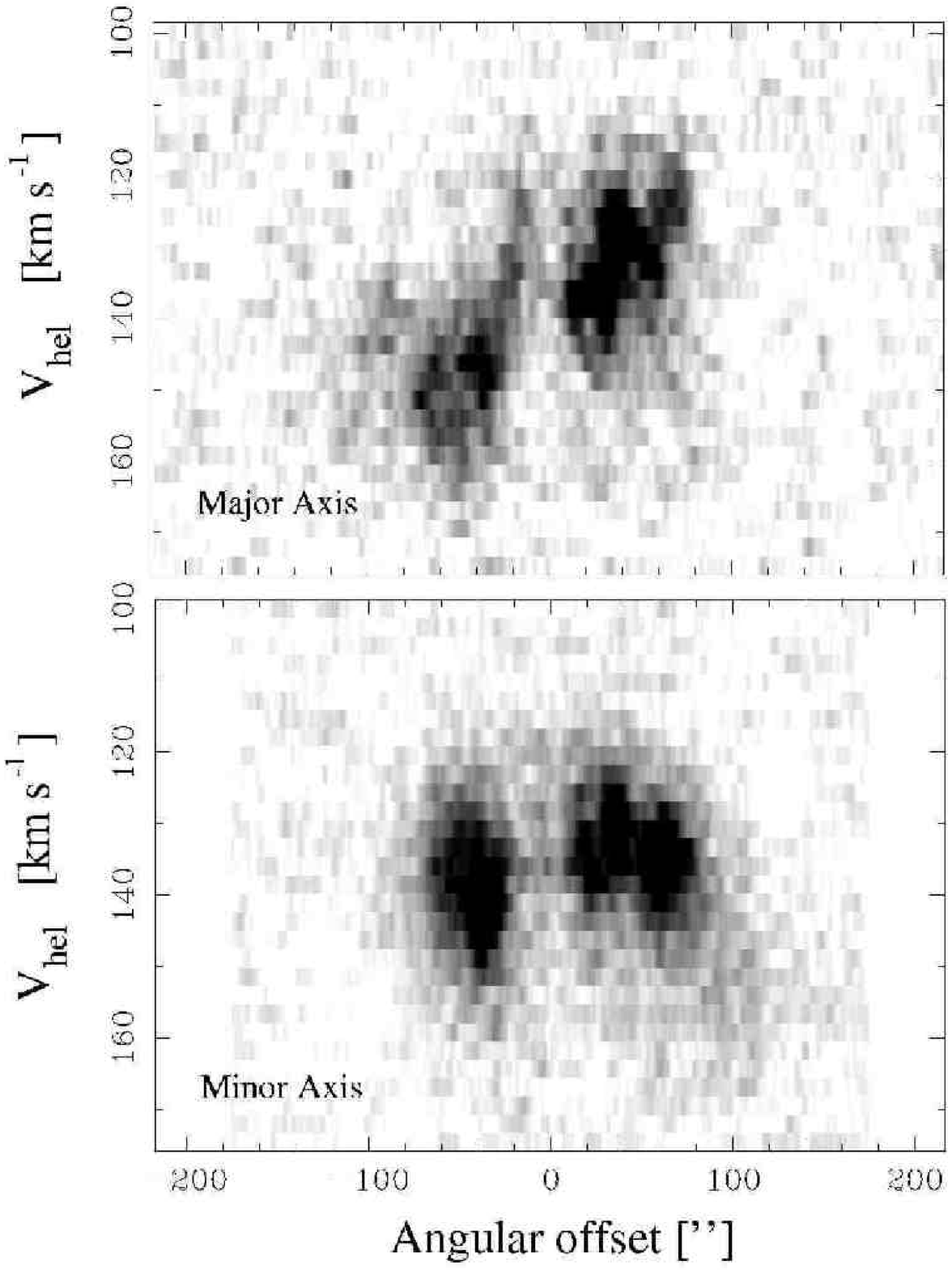}
\figcaption[ott.fig8.eps] {Position--velocity cuts parallel to the
  major and minor axis and centered on the morphological center of
  lowest \ion{H}{1} column density in the hole (cf. Fig.\,\ref{mom0}).
  Positive offsets point to the north--west (major axis) and to the
  south--west (minor axis). 1${\arcmin}$ corresponds to
  $\sim$1\,kpc.\label{pv}}

\plotone{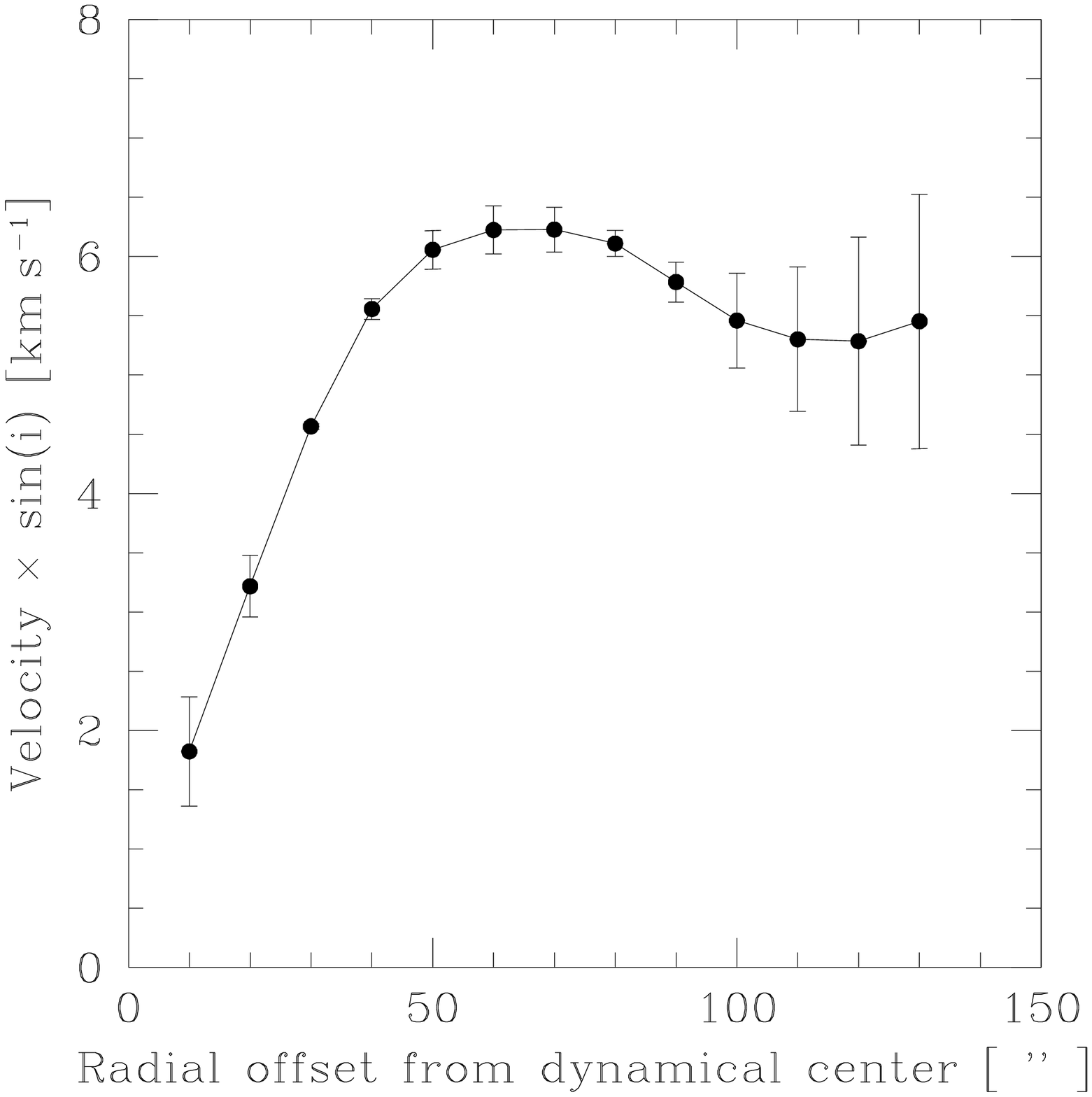}
\figcaption[ott.fig9.eps] {Observed rotational velocity ($V \times \sin i$) curve
of {\holmi}. The error bars reflect the difference in rotational
velocity between the approaching and receding sides of the galaxy.
\label{rotcur}}

\plotone{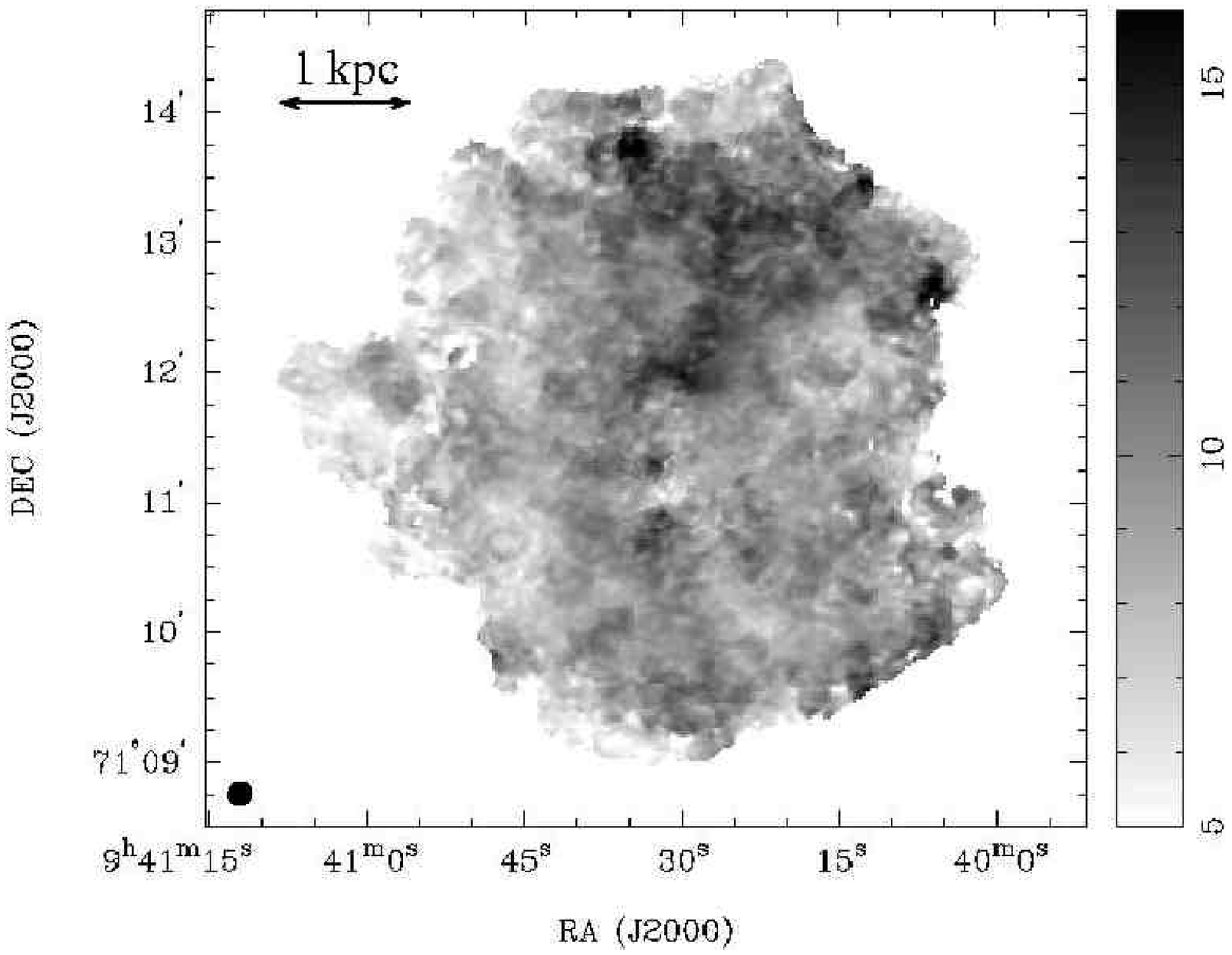}
\figcaption[ott.fig10.eps] {\ion{H}{1} velocity dispersion of
  {\holmi}. In the north--west, the values are systematically higher
  at $\sim 12$\,{\kms} as compared to the rest of the galaxy ($\sim
  9$\,{\kms}). This image was derived using the naturally--weighted
  data which have a resolution of $11\farcs8 \times 11\farcs0$, as
  indicated in the lower left hand corner.
  \label{mom2}}

\plotone{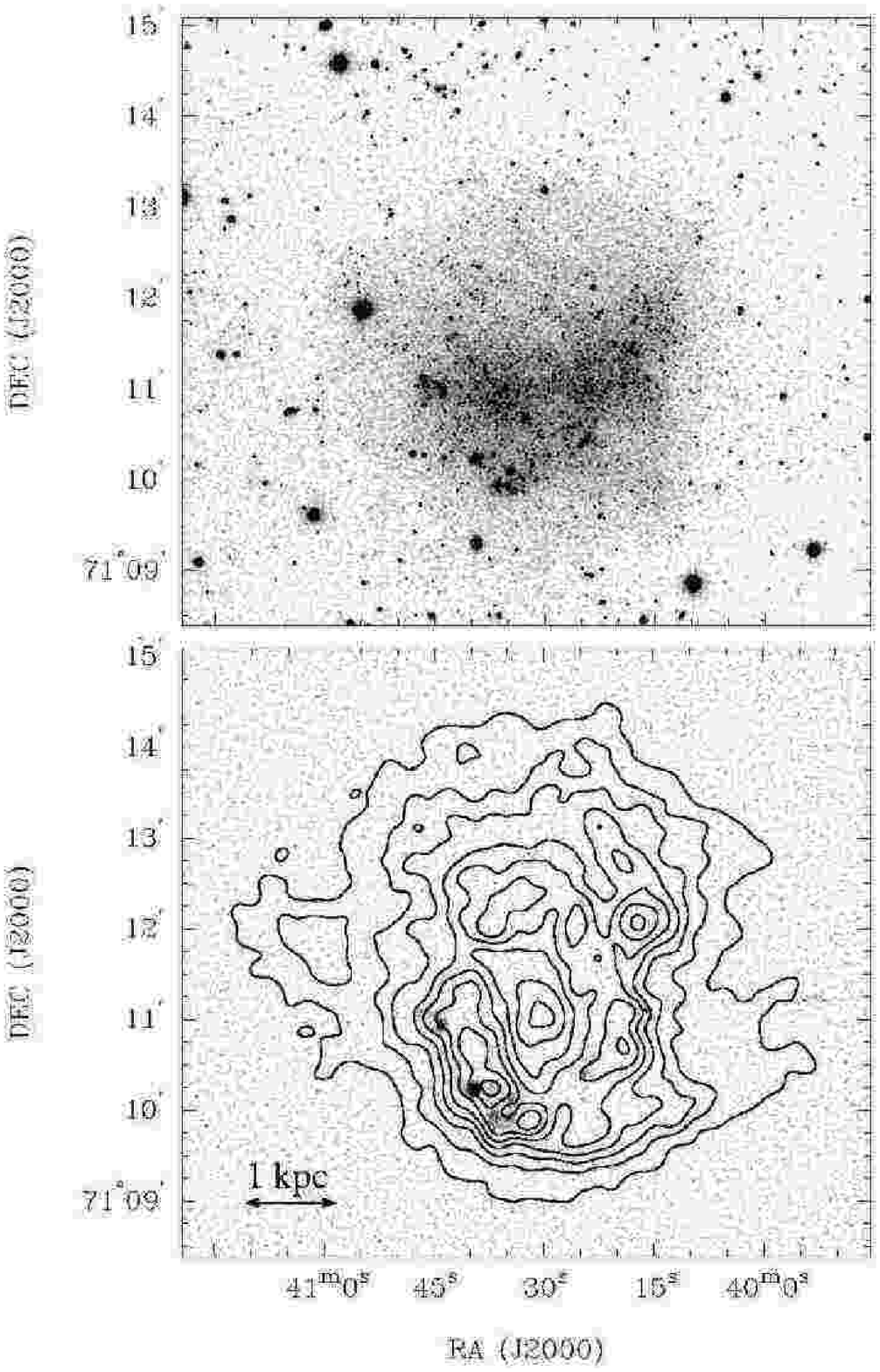}
\figcaption[ott.fig11.eps] {{\it Upper panel}: Optical R$_c$ band
  image of {\holmi}. {\it Lower panel}: Continuum--subtracted
  H$\alpha$ image, at the same scale, with contours of the integrated
  \ion{H}{1} column density smoothed to 15${\arcsec}$. Contour levels
  are drawn at $2\times 10^{20}$\,cm$^{-2}$ intervals, starting at
  $1\times 10^{20}$\,cm$^{-2}$. The optical radiation is mainly
  emitted in the southern region of {\holmi}, within a roughly
  rectangular area which shows an abundance of bright, mainly blue
  stars. The extent over which the R$_c$ band emission and the
  \ion{H}{1} distribution are visible are nearly the
  same.\label{hi+opt}}

\plotone{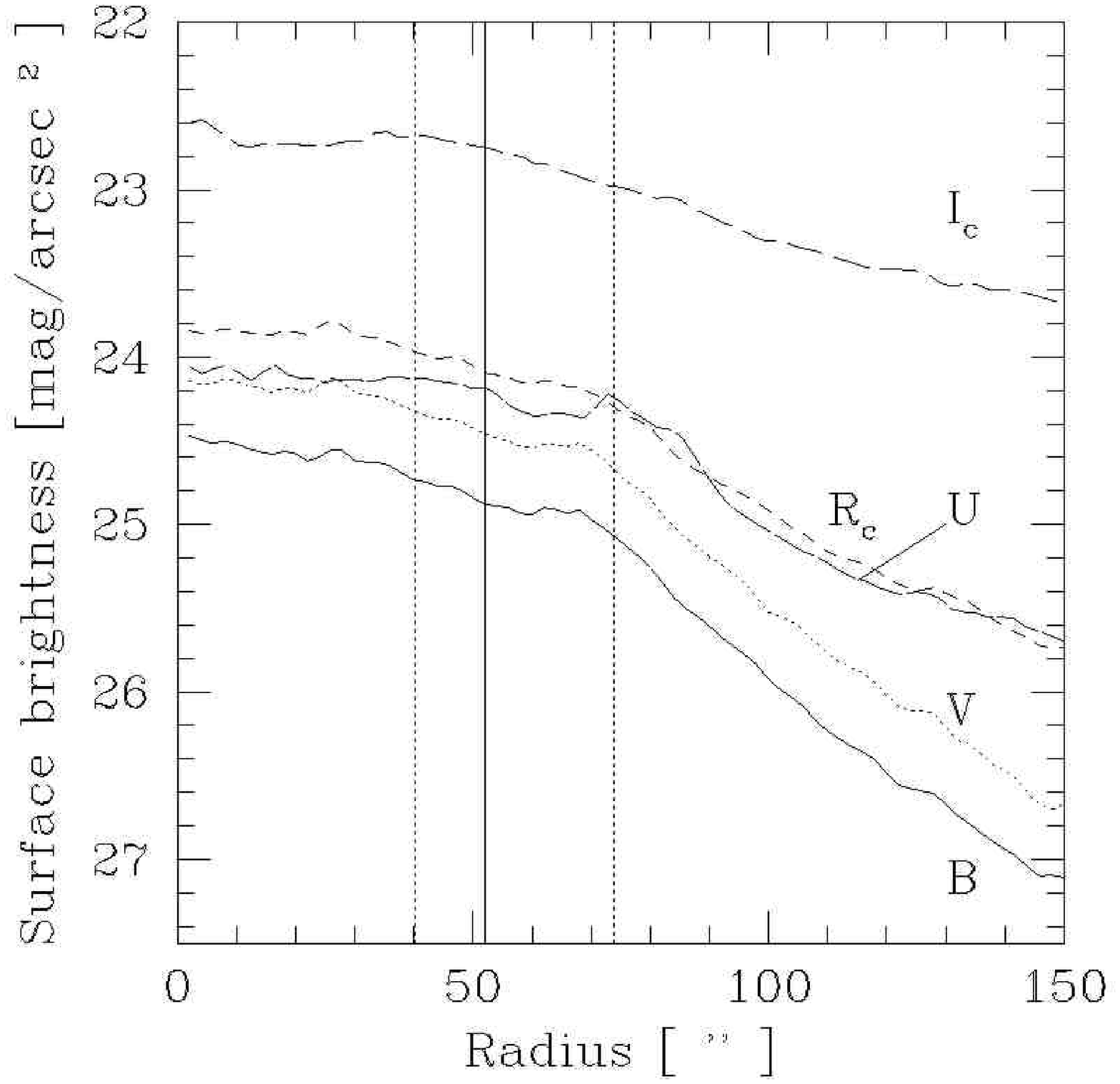}
\figcaption[ott.fig12.eps] {Azimuthally--averaged UBV(RI)$_c$ surface
  brightness distributions with respect to the morphological center of
  the \ion{H}{1} column density distribution (cf. Fig.\,\ref{hirad}).
  At a radius of $\sim 70{\arcsec}$ ($\approx$ 1.2\,kpc) there is a
  change in slope, diminishing in importance with increasing
  wavelength. The vertical lines correspond to the location of the
  maximum surface density of the \ion{H}{1} ring ({\it solid line})
  and the range where the \ion{H}{1} column density exceeds the
  putative star formation threshold, $N_{{\rm HI}}=10^{21}$\,cm$^{-2}$
  ({\it dotted lines}); see also Figs.\,\ref{hirad} and
  \ref{b-r+halpha}. \label{profiles}}

\plotone{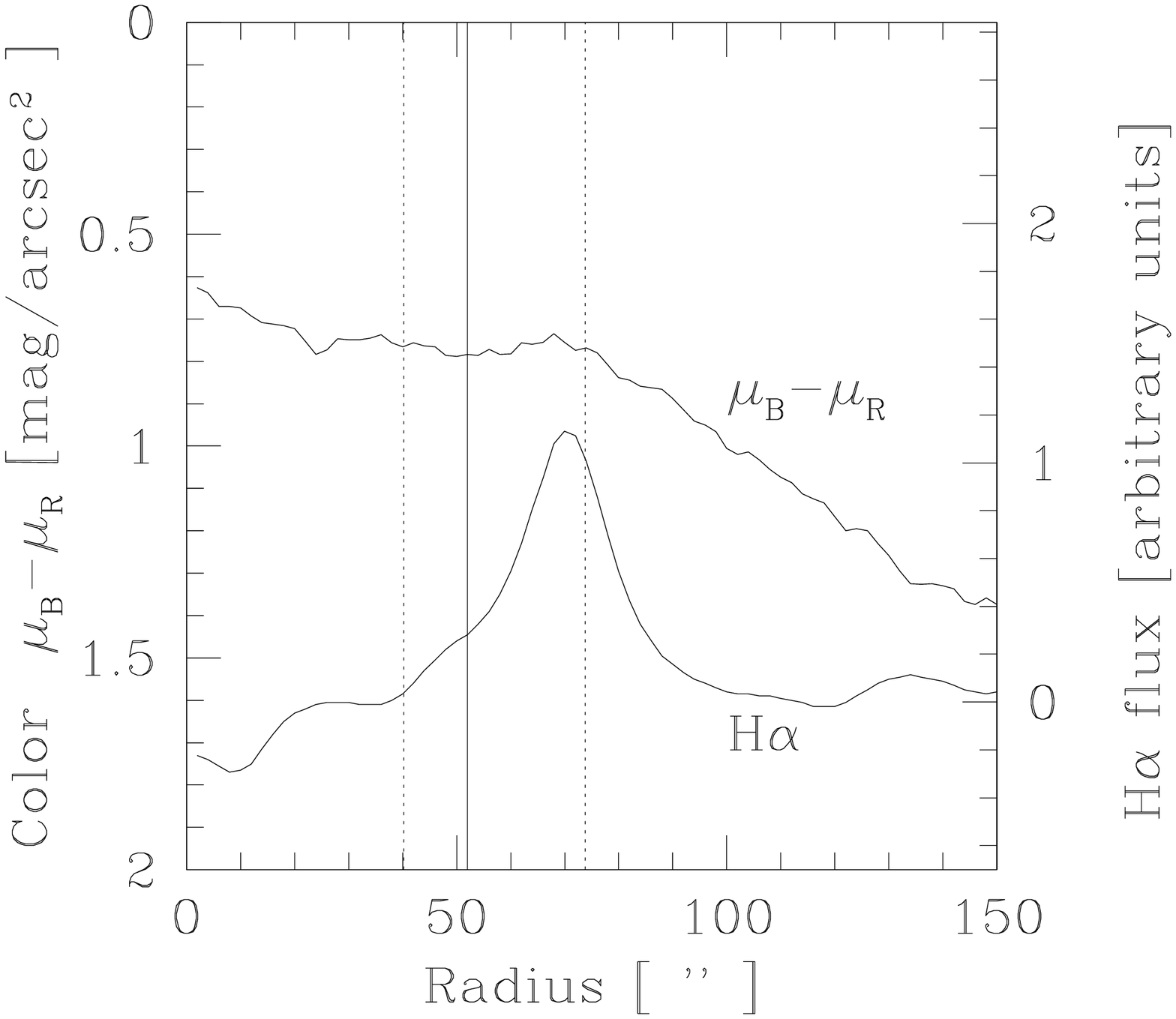}
\figcaption[ott.fig13.eps] {Azimuthally--averaged B$-$R$_c$ color (in
  mag/$\sq\arcsec$, {\it upper graph}) and the smoothed {\Ha} emission
  as a function of radius in arbitrary units ({\it lower graph}). Note
  that the {\Ha} peak and the change in slope in the B$-$R$_c$ graph
  are coincident with the change in slope of all colors at a radius of
  $\sim 70{\arcsec}$ ($\approx$ 1.2\,kpc); see also
  Fig.\,\ref{profiles}. The {\it solid vertical line} shows the peak
  position of the \ion{H}{1} ring, whereas the {\it dotted lines} are
  drawn at the star formation threshold ($N_{{\rm
    HI}}=10^{21}$\,cm$^{-2}$).\label{b-r+halpha}}

\plotone{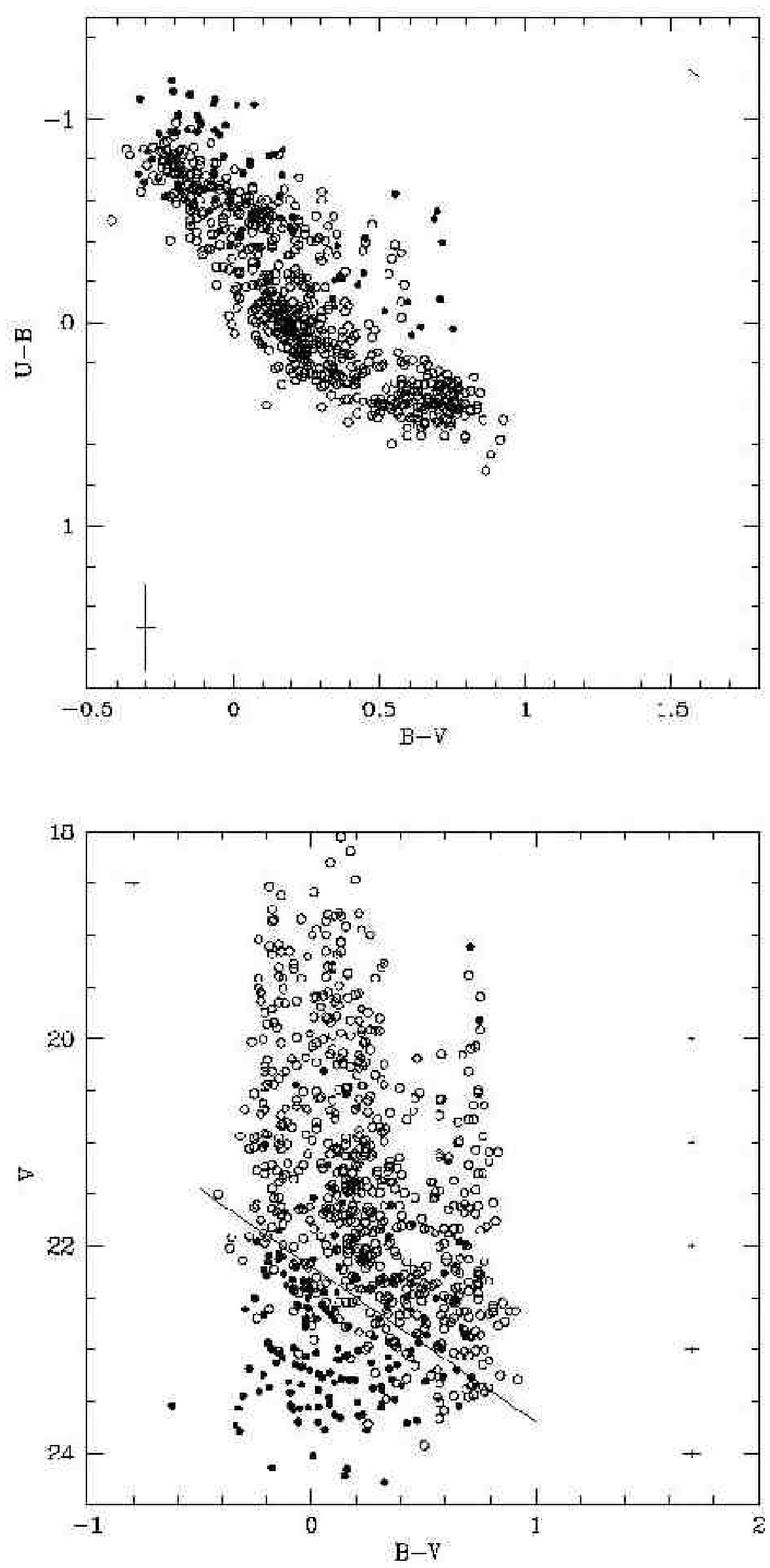}
\figcaption[ott.fig14.eps] {{\it Filled circles}: (U$-$B, B$-$V)
  color--color diagram and (V, B$-$V) color--magnitude diagram of
  {\holmi}. {\it Open circles}: LMC cluster distribution
  \citep{bic96}. A reddening correction corresponding to that towards
  the position of {\holmi} has been applied. The calibration error in
  the color--color diagram is plotted in the lower left. In the same
  diagram, the reddening vector of {\holmi} is shown in the upper
  right. The errors in the CMD are the same as in
  Fig.\,\ref{cmds}.\label{cluster}}

\plotone{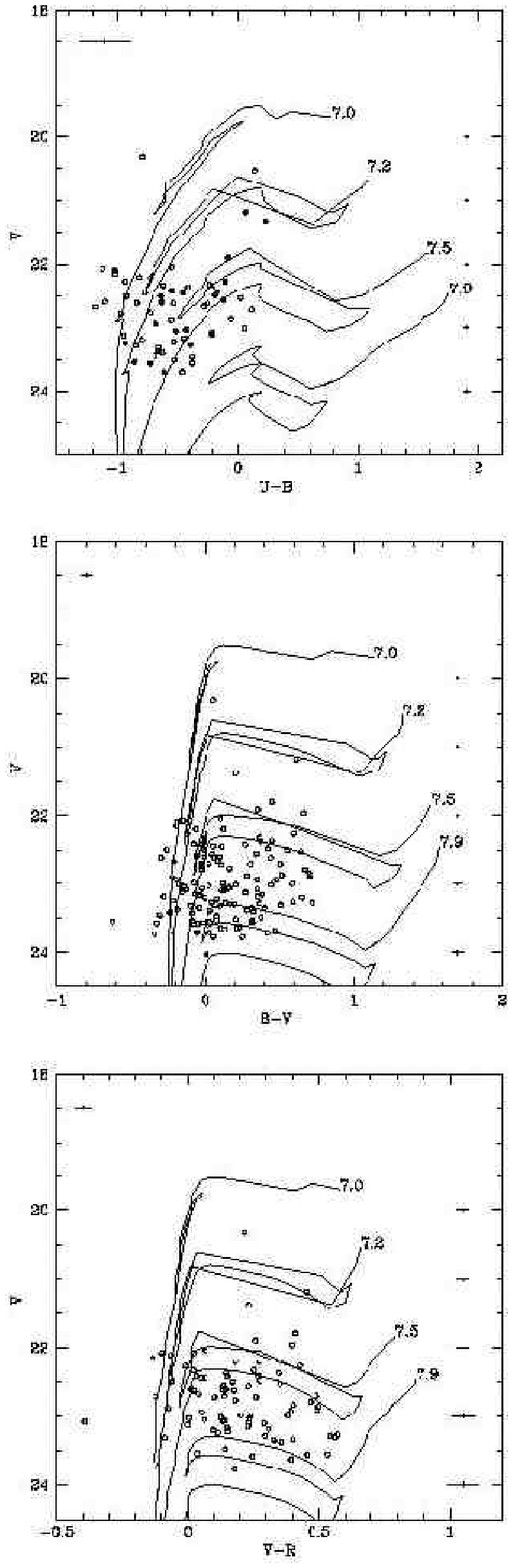}
\figcaption[ott.fig15.eps] {Color--magnitude diagrams of {\holmi}.
  The errors on the right side of the panels are the
  magnitude--dependent formal errors given by the task {\sc allstar}.
  In the upper left corners, the photometric calibration errors are
  shown. Isochrones from \citet{ber94} for a metallicity of Z=0.001
  are overplotted with the logarithmic age in years. These isochrones
  are reddened according to the Galactic foreground values at the
  position of {\holmi}. The {\it filled circles} in the (V, U$-$B)
  diagram are all the stars within a radius of 0\farcm9, \emph{inside}
  the \ion{H}{1} ring. \label{cmds}}

\plotone{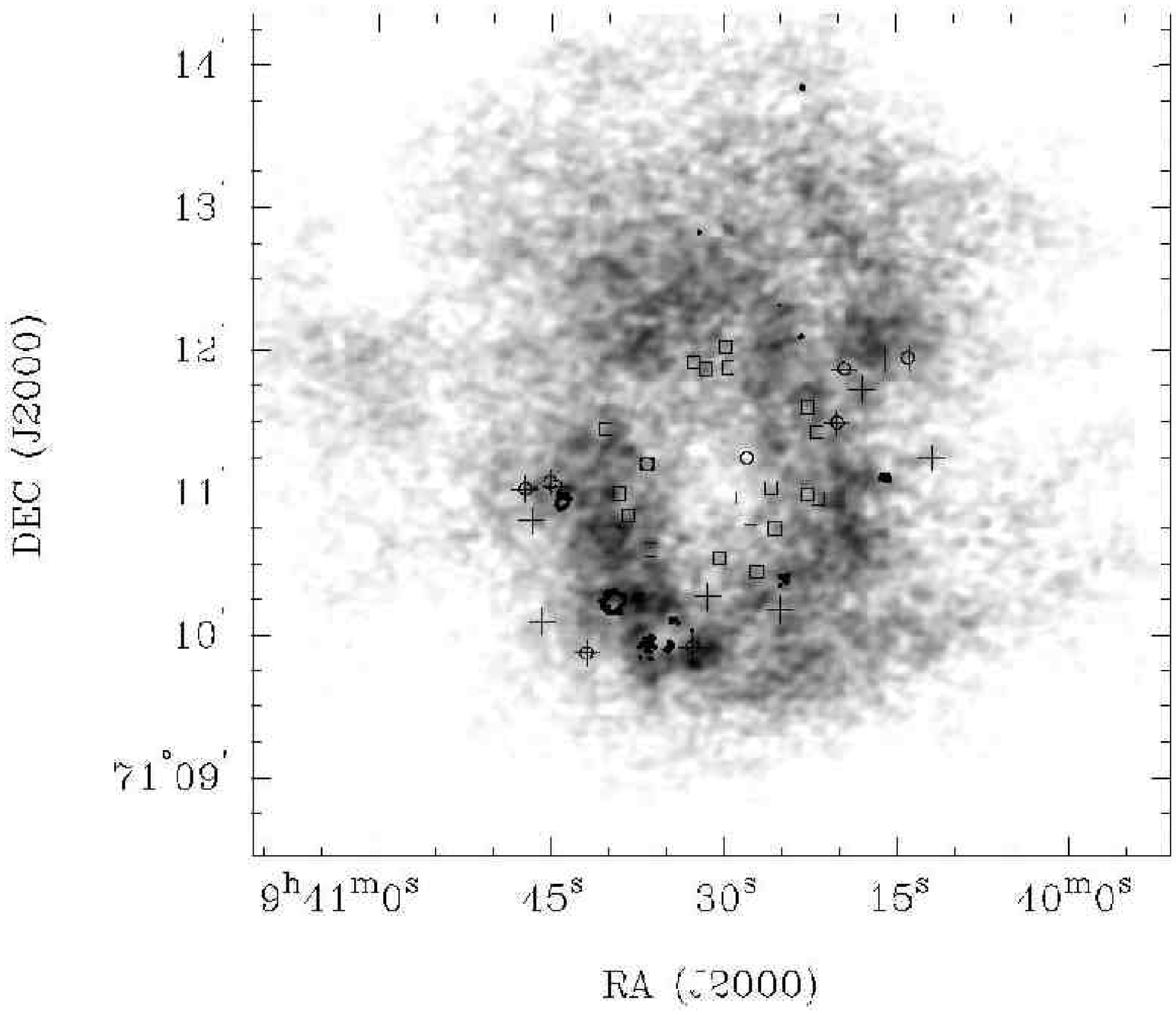}
\figcaption[ott.fig16.eps] {Integrated {\HI} of {\holmi} with {\Ha}
  emission as contours. The markers show the location of young stars
  selected on the basis of the (V, U$-$B) diagram under the condition
  (U$-$B)$\lesssim -0.7$ and V$\lesssim 23$ ({\it crosses}). The {\it
   circles} are stars taken from the (V, B$-$V) diagram with
  (B$-$V)$\lesssim -0.1$ and V$\lesssim 23$. The {\it open squares}
  show the location of the stars plotted in Fig.\,\ref{cmds} as {\it filled
  circles}.\label{histars}}

\plotone{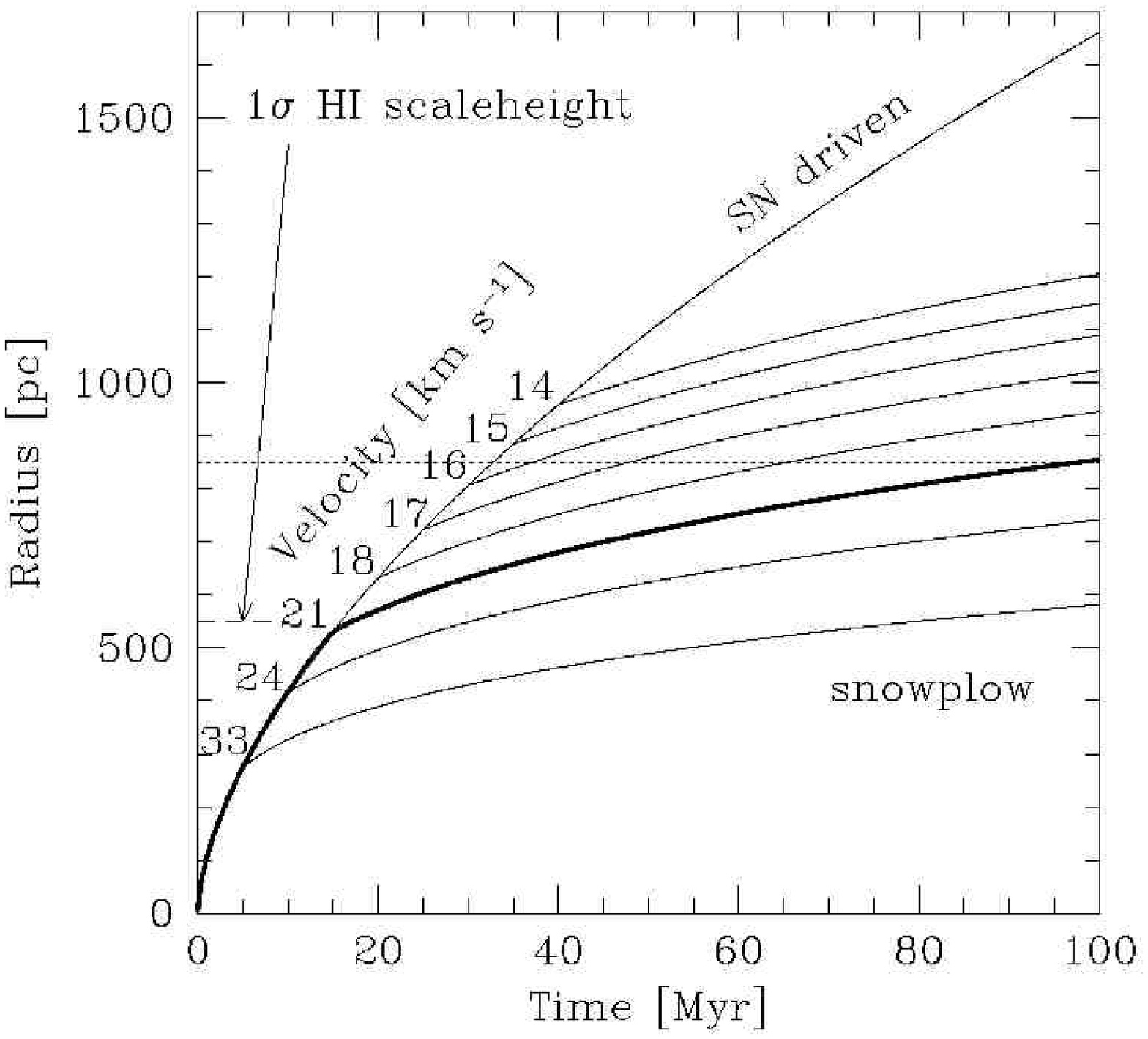}
\figcaption[ott.fig17.eps]{Evolution of the supergiant \ion{H}{1}
  shell which dominates {\holmi} according to the models discussed in
  the text (Sect.\,\ref{creation}). After a certain period of
  SN--driven expansion the shell enters the ``snowplow'' phase, due to
  break--out or cooling. The turn--off points are plotted for critical
  times $t_c$=5--40\,Myr in steps of 5\,Myr. The {\it dotted line}
  indicates the observed present--day radius of the supergiant shell.
  The {\it dashed line} represents the 1$\sigma$ scaleheight of the
  \ion{H}{1} layer of {\holmi}. Expansion velocities, derivatives of
  the radial evolution, are given at the time of the
  break--out.\label{shellsim}}

\plotone{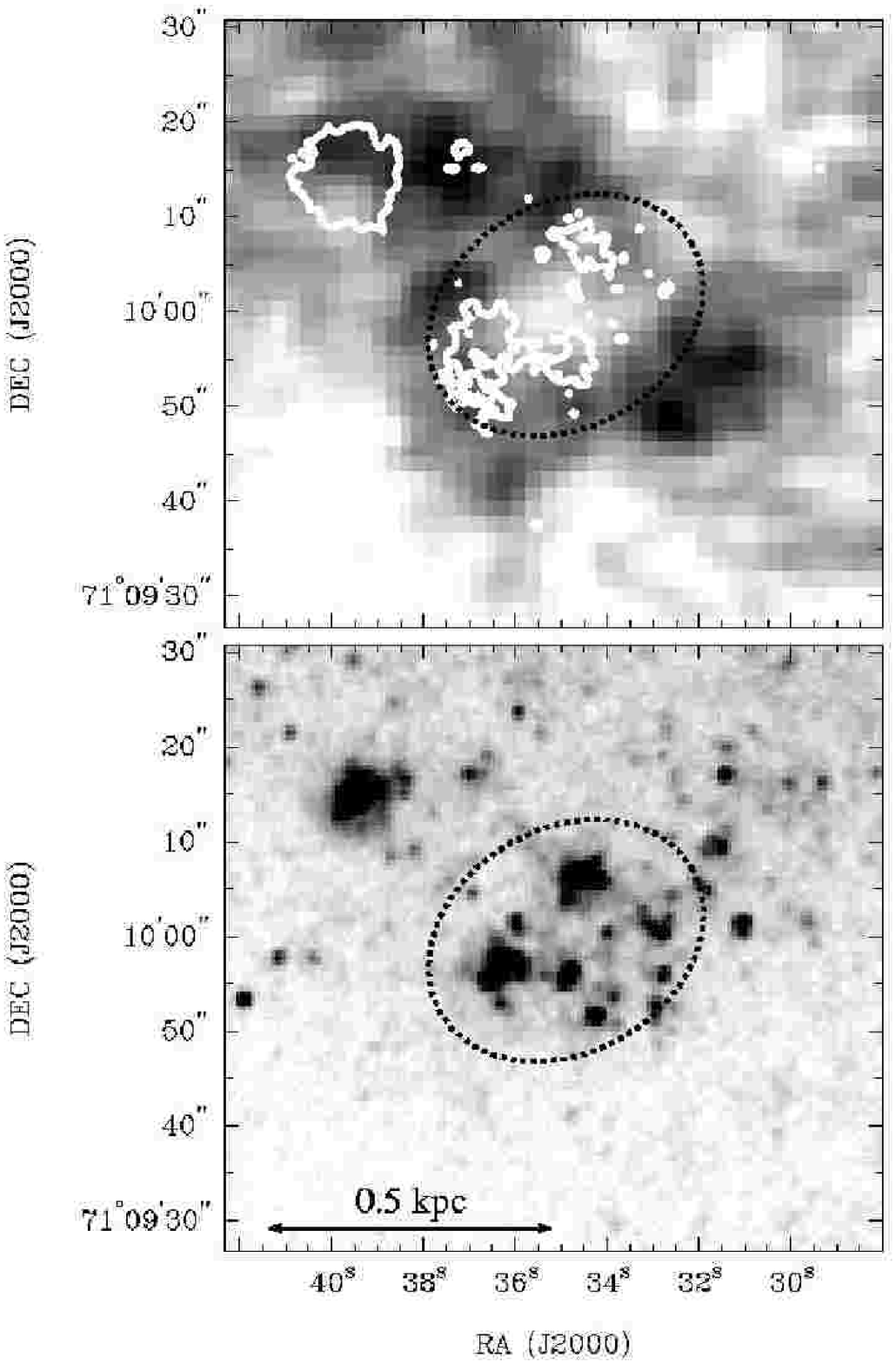}
\figcaption[ott.fig18.eps] {Blow--up of the box marked in
  Fig.\,\ref{mom0}. {\it Upper panel}: {\Ha} emission overlaid as
  white contours on the \ion{H}{1} content. {\it Lower panel}: Johnson
  B image. Both images are at the same scale. The ellipse indicates
  the size and orientation of the \ion{H}{1} shell.\label{region1}}

\plotone{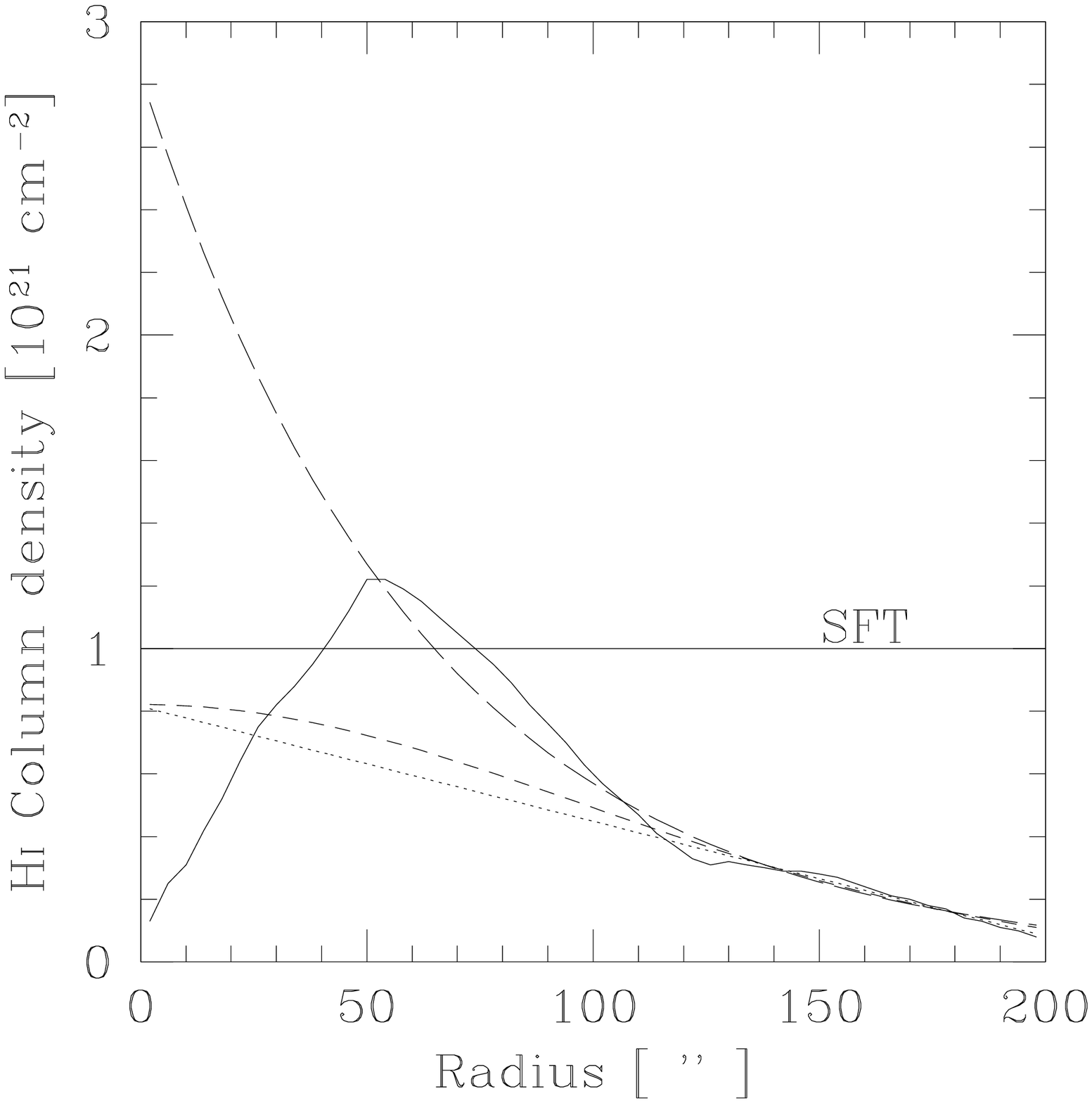}
\figcaption[ott.fig19.eps] {Radial \ion{H}{1} distribution of {\holmi}
  ({\it solid}) with three different extrapolations towards the
  center, in an attempt to estimate the amount of \ion{H}{1} which
  might have been present before the energetic events which created
  the giant hole; exponential ({\it long dashed}), Gaussian ({\it
   short dashed}) and linear ({\it dotted}) extrapolation. The
  horizontal line labelled ``SFT'' displays the empirical SF threshold
  at $N_{\rm HI}=10^{21}$\,cm$^{-2}$.\label{interpol}}

\clearpage

\begin{table}

\begin{tabular}{|l|ccc|}
\tableline
\tableline
                &\multicolumn{3}{c|}{VLA configuration}\\
\tableline
                &B              &C              &D              \\
\tableline
Date of observation&1990 Jul 23&1990 Dec 02&1991 Mar 07 \\
Time on source  &340\,min       &115\,min       &97\,min        \\
Secondary Calibrator    &0917+624       &0836+710       &0836+710       \\
\tableline
Coordinates of pointing center (B1950.0)&\multicolumn{3}{c|}{$\alpha =
09^h$\,36$^m$\,00$^s$, $\delta =
71{\degr}$24${\arcmin}$47${\arcsec}$}\\ 
Primary Calibrator&\multicolumn{3}{c|}{3C286}\\
Number of channels        &\multicolumn{3}{c|}{102}\\
Velocity range  &\multicolumn{3}{c|}{321.39 -- 58.59\,{\kms}}\\
Channel width&\multicolumn{3}{c|}{2.6\,{\kms}}\\
Primary Beam (HPBW)     &\multicolumn{3}{c|}{32${\arcmin}$}\\
Synthesized Beam (HPBW)&\multicolumn{3}{c|}{}\\
\multicolumn{1}{|r|}{``natural'' weighting}&\multicolumn{3}{c|}{11\farcs8$\times$11\farcs0}\\
\multicolumn{1}{|r|}{``uniform'' weighting}&\multicolumn{3}{c|}{6\farcs0$\times$4\farcs7}\\
\multicolumn{1}{|r|}{``robust'' weighting}&\multicolumn{3}{c|}{8\farcs2$\times$7\farcs0}\\
1$\sigma$ rms per channel map (``robust'' weighting)&\multicolumn{3}{c|}{1.40\,mJy\,beam$^{-1}$}\\

\tableline
\tableline
\end{tabular}

\caption{Properties of the VLA \ion{H}{1} observations.\label{tabhiprop}}

\end{table}



\begin{table}

\begin{tabular}{|l|ccccc|}

\tableline\tableline
Johnson (Cousins) Band&U&B&V&R$_c$&I$_c$\\
\tableline
Apparent magnitude [mag]&\phantom{$-$}13.23&\phantom{$-$}13.25&\phantom{$-$}12.49&\phantom{$-$}12.40&11.87\\
Absolute magnitude [mag]&$-14.57$&$-14.55$&$-15.31$&$-15.40$&$-15.93$\\
Mean surface brightness [mag/$\sq {\arcsec}$]&\phantom{$-$}25.35&\phantom{$-$}25.37&\phantom{$-$}24.61&\phantom{$-$}24.52&\phantom{$-$}23.99\\
Central surface brightness [mag/$\sq {\arcsec}$]&\phantom{$-$}24.06&\phantom{$-$}24.47&\phantom{$-$}24.15&\phantom{$-$}23.84&\phantom{$-$}22.60\\
\tableline
Blue luminosity [$L_{{\rm B}_\sun}$]&\multicolumn{5}{c|}{$1.0\times10^8$}\\
${\cal M}_{\rm HI}/ {L_{\rm B}}$ [${\cal M}_\sun/ {L_{{\rm B}_{\sun}}}$]&\multicolumn{5}{c|}{1.1}\\
\tableline\tableline
\end{tabular}

\caption{General optical properties of {\holmi}. The mean surface brightness is based on 
a circular area with a radius of 150\arcsec.\label{opttable}}

\end{table}


\begin{table}

\renewcommand{\arraystretch}{0.8}
\begin{tabular}{|l|c|}

\tableline\tableline
Object Name&{\holmi} (UGC\,5139, DDO\,63)\\
\tableline\tableline
Position (J2000)&$\alpha = 09^{h}40^{m}32\fs3, \delta = +71{\degr}10{\arcmin}56{\arcsec}$\\
Morphological type\tablenotemark{a}&IAB(s)m\\
Adopted distance&3.6\,Mpc (m$-$M=27.80\,mag)\\
Scale at this distance& 1{\arcmin} $\hat{\approx}\ 1$\,kpc\\
\ion{H}{1} diameter&5.8\,kpc\\
Systemic velocity&141.5\,{\kms}\\
Inclination&$10{\degr}\lesssim i \lesssim 14{\degr}$\\
\ion{H}{1} linewidth (FWHM)&27.1\,\kms\\
Total \ion{H}{1} flux&$36.0\pm4.0$\,Jy\,{\kms}\\
Total \ion{H}{1} mass&1.1$\times 10^{8}\,{\cal M}_\sun$\\
Mean \ion{H}{1} column density&3.9$\times 10^{20}$\,cm$^{-2}$\\
Peak \ion{H}{1} column density&2.0$\times 10^{21}$\,cm$^{-2}$\\
Average midplane \ion{H}{1} particle volume density&0.10\,cm$^{-3} \lesssim n_0 \lesssim 0.20\,$cm$^{-3}$\\
Average \ion{H}{1} 1$\sigma$ scaleheight&$250\,{\rm pc}\lesssim h \lesssim 550$\,pc\\
Mean \ion{H}{1} velocity dispersion&9\,{\kms}\\
Total mass&$\lesssim 5.5 \times 10^{8}\,{\cal M}_{\sun}$\\
Dark matter content&$\lesssim 3.1 \times 10^{8}\,{\cal M}_{\sun}$\\
Dynamical center (J2000)&$\alpha = 09^h 40^m 31.6^s$, $\delta
= 71{\degr} 11{\arcmin} 45{\arcsec}$\\
Morphological center (J2000)&$\alpha = 9^h 40^m 30^s$, $\delta  = 71{\degr} 11{\arcmin} 1\farcs8$\\
Optical diameter at $\mu_{\rm B}=25\,mag/\sq{\arcsec}$&3.7\,kpc $\times$ 1\,kpc\\
Adopted extinction\tablenotemark{b}&E$_{\rm B-V}$=0.05\,mag\\
Apparent blue luminosity&13.25\,mag\\
Absolute blue luminosity&$-14.55\,{\rm mag}\ \hat{=}\ 1.0 \times 10^8 L_{{\rm B}_{\sun}}$\\
Central blue surface brightness&24.47\,mag/\sq{\arcsec}\\
Mean blue surface brightness&25.69\,mag/\sq \arcsec\\
Oxygen abundance\tablenotemark{c} $12+{\rm log}(\frac{{\rm O}}{{\rm H}})$&7.7\\
Current star formation rate\tablenotemark{d}&0.004\,${\cal M}_\sun$\,yr$^{-1}$\\
\tableline\tableline
\end{tabular}

\tablenotetext{a}{from the Nearby Extragalactic Database (NED)}
\defcitealias{sch98}{Schlegel et al. (1998)}
\tablenotetext{b}{\citetalias{sch98}}
\tablenotetext{c}{\citet{mil96}}
\tablenotetext{d}{\citet{mil94}}

\caption{Summarized general properties of {\holmi}.\label{generaltable}}

\end{table}


\begin{table}

\renewcommand{\arraystretch}{1}
\begin{tabular}{|l|c|}

\tableline\tableline
Radius&0.85\,kpc\\
FWHM width&1.26\,kpc\\
\ion{H}{1} mass&$\sim 8\times 10^{7}\,{\cal M}_\sun$\\
Central \ion{H}{1} column density&$6\times 10^{19}$\,cm$^{-2}$\\
Average \ion{H}{1} column density on the rim&$1.2\times 10^{21}$\,cm$^{-2}$\\
Peak \ion{H}{1} column density on the rim&$2.0\times 10^{21}$\,cm$^{-2}$\\
Age&$80\pm 20$\,Myr\\
Energy&$1.2\times 10^{53}$\,erg ($\hat{=}$ 120 SNe) $\lesssim E$\\
&$ \lesssim 2.6 \times 10^{53}$\, erg ($\hat{=}$ 260 SNe)\\
\tableline\tableline
\end{tabular}

\caption{Properties of the central supergiant \ion{H}{1} shell.\label{ringtable}}

\end{table}

\clearpage


\end{document}